\documentclass[letterpaper,journal]{IEEEtran}
\usepackage{hyphenat}

\usepackage{amsmath,amsfonts}
\usepackage{algorithmic}
\usepackage{array}
\usepackage[caption=false,font=normalsize,labelfont=sf,textfont=sf]{subfig}
\usepackage{textcomp}
\usepackage{stfloats}
\usepackage{url}
\usepackage{verbatim}
\usepackage{graphicx}
\usepackage{bm}
\usepackage{booktabs}
\usepackage{arydshln}
\usepackage{nicematrix}
\usepackage{siunitx}
\usepackage{hyperref}

\usepackage{tikz}
\usetikzlibrary{positioning, calc, arrows.meta}

\usepackage{microtype}

\def\BibTeX{{\rm B\kern-.05em{\sc i\kern-.025em b}\kern-.08em
    T\kern-.1667em\lower.7ex\hbox{E}\kern-.125emX}}
\usepackage{balance}

\newcommand{\ssub}[1]{_{\scriptscriptstyle #1}}
\DeclareSIUnit\samples{S}
\providecommand{\GSs}{GS/s}
\providecommand{\cfg}[2]{$#1{\times}#2$}

\newcommand{\Ag}{\bm{\mathsf{A}}}                     
\newcommand{\Bg}{\bm{\mathsf{B}}}                     
\newcommand{\Agm}{\bm{\mathsf{A}}_{-}}                
\newcommand{\Bgm}{\bm{\mathsf{B}}_{-}}                

\newcommand{\yb}{\bm{y}}                      
\newcommand{\xb}{\bm{x}}                      
\newcommand{\xg}{\bm{\mathsf{x}}}            
\newcommand{\yg}{\bm{\mathsf{y}}}            
\newcommand{\xbp}{\bar{\bm{x}}}               
\newcommand{\ybp}{\bar{\bm{y}}}               

\newcommand{\SR}{\bm{R}}               

\newcommand{\vg}{\bm{\mathsf{v}}}            
\newcommand{\wg}{\bm{\mathsf{w}}}            
\newcommand{\vbp}{\bar{\bm{v}}}               
\newcommand{\wbp}{\bar{\bm{w}}}               

\newcommand{\ye}{\bm{y_e}}
\newcommand{\we}{\bm{w_e}}

\begin{document}
\bstctlcite{IEEEexample:BSTcontrol}

\title{Parallel Cascaded Recursive Filtering on Multi-Core CPUs and GPUs}

\author{Haotian~Zhai,~\IEEEmembership{Student~Member,~IEEE,}
        and~Bernd-Peter~Paris,~\IEEEmembership{Senior~Member,~IEEE}%
\thanks{This work was supported in part by the National Science Foundation under grant 2029836.
H.~Zhai and B.-P.~Paris are with the Department of Electrical
and Computer Engineering, George Mason University, Fairfax, VA 22030
USA (e-mail: hzhai@gmu.edu; pparis@gmu.edu).}}


\maketitle

\begin{abstract}
The companion paper of this two-part series reformulated cascaded
second-order (biquad) recursive filtering as a block-tridiagonal
linear system and developed two parallel solution algorithms, PH
factorization and cyclic reduction, reaching over
600~Megasamples per second on a single SIMD core. This paper scales
that framework to multi-core CPUs and GPUs, where a new obstacle
appears: the terminal outputs of each signal block group are the
initial conditions of the next, so naively distributed groups
serialize. The dependency is resolved by superposition---each
group's output splits into a zero-state response, computable
immediately, and a homogeneous correction applied when the state
arrives---and by a divide-and-conquer form of cyclic reduction that
exposes both terminal blocks before back substitution, as
asynchronous state propagation requires. Two implementations pair
the two dominant deployment scenarios with opposite treatments of
the dependency. For real-time streaming, a wavefront pipeline
realized with TBB flow graphs parallelizes across cascade sections,
preserves first-in-first-out order, and achieves $3.95\times$
scaling on six performance cores, about 2.4~Gigasamples per second
for a 16th-order filter. For batched processing, a single-kernel GPU
implementation carries each group through the entire cascade in
registers and parallelizes across groups with a decoupled lookback
protocol; a communication-based cost model, comprising a memory
roof, a barrier price, and a latency-hiding floor, reduces tuning to
two parameters and predicts the measured behavior across two GPU
generations. The best kernels reach 38.2~Gigasamples per second for a single second-order section on
an RTX~3060, 85\% of the memory-bandwidth roof, exceed the strongest
published parallel recurrence baseline at every filter order, and
remain numerically valid at order 16, where the direct-form baseline
fails.
\end{abstract}


\section{Introduction}
\label{sec:intro}

\IEEEPARstart{R}{ecursive} filters are among the most fundamental
building blocks in digital signal processing. Compared to
non-recursive (FIR) filters, they typically require far fewer
coefficients to meet a given magnitude-response specification,
translating directly into fewer computations per unit time. This
efficiency makes them especially attractive in high-throughput
settings such as real-time digital filtering in modern communication
receivers~\cite{Valkama_01}, speech and audio
processing~\cite{Belloch_14}, high-resolution image
restoration~\cite{Nehab_11}, and video processing~\cite{Tammana_23}.
However, the feedback dependency in a recursive filter, where each
output sample depends on previous outputs, creates a sequential
bottleneck that limits computational speed. Approaches to this
bottleneck span half a century, from FFT-based approximations of the
impulse response~\cite{Helms_67,Voelcker_70} through block
filtering~\cite{Burrus_71,Burrus_72} to SIMD implementations on
vector DSP architectures and commodity
CPUs~\cite{Sung_92,Kutil_08,Ahn_09,Lee_10,Zhai_24}. In the first
part of this two-part series~\cite{Zhai_26}, we addressed it for
the most practically relevant realization---the cascade of second-order
sections---through a block-matrix reformulation, \emph{multi-block
filtering}, in which a stride-$N$ permutation maps a group of $NL$
samples into a block-tridiagonal system solved by two parallel
algorithms, PH factorization and cyclic reduction. On a single SIMD
core, the resulting implementation reaches
\SI{616}{\mega\samples\per\second} for a 16th-order filter, an
$8\times$ improvement over the standard
\texttt{scipy.signal.sosfilt} routine~\cite{virtanen_20}.

A single core, however, is no longer where high-throughput
computing resides: commodity CPUs integrate a dozen or more heterogeneous
cores, and GPUs keep tens of thousands of threads resident to hide
latency~\cite{Volkov_16}. Scaling multi-block filtering across such
units meets an obstacle that does not exist within one core: the
\emph{inter-group state dependency}. The last two outputs of each
signal block group are the initial conditions of the next group in
every section \cite{Zhai_26}, so a naive distribution of groups across processing
units serializes them. Moreover, two different deployment
scenarios impose different disciplines: real-time streaming, as in a
software-defined radio receive chain, requires low latency and
first-in-first-out output order, while batched processing of
recorded data requires maximum throughput and imposes no order. The two scenarios favor different architectures, and, as this
paper shows, different algorithmic treatments of the same
dependency.

Prior work offers partial answers. Building on the classical
prefix-scan constructions~\cite{Blelloch_96,jaja_92}, the modern
single-pass GPU scan resolves the analogous dependency of
first-order recurrences through \emph{decoupled lookback}, in which
each block publishes a local result and inspects its predecessors'
published states~\cite{Merrill_16}. Maleki et~al.\ extended
scan-based evaluation to general $m$th-order linear
recurrences~\cite{Maleki_18}, but in the direct-form realization,
whose per-thread state grows with the order and whose coefficients
are numerically fragile for high-order
filters~\cite{Oppenheim_10}. GPU implementations of block IIR
filtering launch separate kernels per computation stage, with every
intermediate result passing through global memory~\cite{Lee_13}, and
GPU recursive filtering for image processing has relied on
overlapped blocks tailored to the two-dimensional
setting~\cite{Nehab_11}. On CPUs, task-based runtimes such as
Threading Building Blocks provide pipelined parallelism across
cores~\cite{Michael_19}, but published parallel IIR work targets the
direct form rather than the cascade. No existing work, to our
knowledge, integrates inter-group state propagation with group-level
parallel filtering algorithms for the cascaded second-order
realization on either architecture.

This paper fills that gap. We extend the multi-block filtering
framework of the first part~\cite{Zhai_26} to multi-core CPUs and GPUs,
covering both deployment scenarios. The main contributions are as
follows.

\begin{enumerate}
\item \textbf{Inter-group state propagation:} We derive the
  state-bearing forms of PH factorization and cyclic reduction: by
  superposition, each group's output splits into a zero-state
  response, computable before the group's state is known, and a
  homogeneous correction applied when the state arrives, allowing
  groups to execute concurrently.

\item \textbf{Divide-and-conquer cyclic reduction:} We reformulate
  cyclic reduction so that both terminal blocks of a group are
  available before back substitution---the property that
  asynchronous state propagation requires.

\item \textbf{Real-time multi-core CPU implementation:} A wavefront
  pipeline realized with TBB flow graphs preserves FIFO order and
  achieves $3.95\times$ scaling on the six performance cores of a
  heterogeneous Meteor Lake processor,
  \SI{2.4}{\giga\samples\per\second} for a 16th-order filter.

\item \textbf{Batched GPU implementation:} A single kernel carries
  each group through the entire cascade in registers, resolving the
  state dependency with decoupled lookback; a communication-based
  cost model reduces tuning to two parameters and predicts the
  measured behavior across two GPU generations. The best kernels
  reach \SI{38.2}{\giga\samples\per\second} for a single second-order section on an RTX~3060, 85\% of
  the memory-bandwidth roof.

\item \textbf{Cross-architecture evaluation:} The cascade algorithms
  exceed the strongest published GPU recurrence
  engine~\cite{Maleki_18} at every filter order and remain
  numerically valid at order 16, where the direct form fails.
\end{enumerate}

This paper is organized as follows. Section~\ref{sec:background}
reviews the multi-block filtering framework of Part~I and extends it
with the two ingredients that parallel execution requires: the
divide-and-conquer form of cyclic reduction and the inter-group
state propagation algebra. Section~\ref{sec:cpu} describes the
real-time multi-core CPU implementation: wavefront execution, its
realization with TBB flow graphs, and measurements on a
heterogeneous Meteor Lake processor. Section~\ref{sec:gpu} describes
the batched GPU implementation: the thread-block mapping, the
decoupled lookback protocol, a communication-based cost model, the
four algorithms, and measurements on two GPU generations.
Significant throughput improvements over the strongest published
parallel recurrence baseline are shown.
Section~\ref{sec:conclusion} presents a summary and conclusion as
well as future work.

\section{Arithmetic Background}
\label{sec:background}

\noindent This section summarizes the multi-block filtering framework
developed in Part~I~\cite{Zhai_26} and extends it with the two
ingredients that parallel group execution requires: the
divide-and-conquer formulation of cyclic reduction and the principle
of inter-group state propagation.

\subsection{The Cascaded Recursive Filter}
\label{sec:bg:cascade}

A general recursive filter of order $2K$ is realized as a cascade of
$K$ second-order sections (biquads). Each biquad computes
\begin{equation}
\label{eq:biquad}
y[n] = x[n] + b\ssub{1}x[n{-}1] + b\ssub{2}x[n{-}2]
     - a\ssub{1}y[n{-}1] - a\ssub{2}y[n{-}2],
\end{equation}
where the scaling coefficients $b\ssub{0,k}$ of all sections are
split off into a common gain factor. The output of section $k$ is the
input of section $k{+}1$. The cascaded realization confines the
effect of coefficient quantization to a single pole pair per section
and is numerically more robust than the direct
form.

\subsection{Multi-Block Filtering}
\label{sec:bg:multiblock}

The input sequence is partitioned into blocks of $L$ samples, and $N$
consecutive blocks form a \emph{signal block group} of $NL$
samples. Within a group, the samples are permuted with
stride $N$: the $m$-th permuted block is $\xbp\ssub{m} = [x[m],\,
x[m{+}N],\, \ldots,\, x[m{+}(L{-}1)N]]^\top$. Multi-block filtering
describes the action of a single second-order section~\eqref{eq:biquad}
on a signal block group: as shown in~\cite{Zhai_26}, the permuted group
satisfies the linear system
\begin{equation}
\label{eq:multiblock_system}
\Ag\,\yg = \Bg\,\xg + \Bgm\,\xb\ssub{-1} - \Agm\,\yb\ssub{-1},
\end{equation}
where $\xg = [\xbp\ssub{0}, \ldots, \xbp\ssub{N-1}]^\top$, $\yg$ is
defined analogously, and $\xb\ssub{-1} = [x[-2],\, x[-1]]^\top$,
$\yb\ssub{-1} = [y[-2],\, y[-1]]^\top$ are the last two input and
output samples of the preceding group. The matrix $\Ag \in
\mathbb{R}^{NL \times NL}$ is block-Toeplitz with corner coupling,
\begin{equation}
\label{eq:multi_block_A}
\Ag =
\begin{bmatrix}
\bm{I}          &                 &                 & a\ssub{2}\SR    & a\ssub{1}\SR \\
a\ssub{1}\bm{I} & \bm{I}          &                 &                 & a\ssub{2}\SR \\
a\ssub{2}\bm{I} & a\ssub{1}\bm{I} & \bm{I}          &                 &              \\
                & \ddots          & \ddots          & \ddots          &              \\
                &                 & a\ssub{2}\bm{I} & a\ssub{1}\bm{I} & \bm{I}
\end{bmatrix},
\end{equation}
where every $L \times L$ entry is a scalar multiple of the identity
matrix $\bm{I}$ or the lower shift matrix $\SR$. The matrix $\Bg$ has
the same structure with $a\ssub{1}, a\ssub{2}$ replaced by
$b\ssub{1}, b\ssub{2}$, and $\Agm, \Bgm \in \mathbb{R}^{NL \times 2}$
couple the initial state to the first blocks.

The computation of \eqref{eq:multiblock_system} can be separated into two stages. The non-recursive stage
evaluates 
\begin{equation}
\label{eq:feedforward}
\vg = \Bg\,\xg + \Bgm\,\xb\ssub{-1} 
\end{equation}
in parallel across all $N$ blocks. The recursive stage complements $\Agm\,\yb\ssub{-1}$ and solves
\begin{equation}
\label{eq:recursive}
\Ag\,\yg = \vg - \Agm\,\yb\ssub{-1}.
\end{equation}
Note that this differs slightly from the theory paper, where the
initial-state term $\Agm\,\yb\ssub{-1}$ was folded into the
right-hand side of \eqref{eq:feedforward}. The reason is that
the theory paper focuses on single-core execution: serial group
execution guarantees that $\yb\ssub{-1}$ is available when a group
starts. This paper targets parallel group execution on multi-core
CPUs and GPUs, and $\yb\ssub{-1}$ holds the last two output samples of
the preceding group. This means that $\Agm\,\yb\ssub{-1}$ carries
the inter-group dependency.

The non-recursive stage is computed trivially in parallel; the
recursive stage is the subject of the remainder of this section.

\subsection{PH Factorization}
\label{sec:bg:ph}

The PH factorization $\Ag = \bm{P}\bm{H}$ decomposes $\Ag$ into a
banded block-Toeplitz factor $\bm{P}$ without corner coupling and a
sparse factor $\bm{H}$ \big[\!\cite{Zhai_26}, Eq.~(23)\big]. Substituting
$\Ag = \bm{P}\bm{H}$ into~\eqref{eq:recursive} and moving
$\Agm\,\yb\ssub{-1}$ to the left-hand side yields
\begin{equation}
\label{eq:recursive_ph_form}
\bm{P}\bigl(\bm{H}\,\yg + \bm{P}^{-1}\Agm\,\yb\ssub{-1}\bigr) = \vg.
\end{equation}
Defining
\begin{equation}
\label{eq:PH_w_def}
\wg = \bm{H}\,\yg + \bm{P}^{-1}\Agm\,\yb\ssub{-1},
\end{equation}
the solution proceeds in two stages: first solve
\begin{equation}
\label{eq:PH_PART}
\bm{P}\,\wg = \vg
\end{equation}
for $\wg$, then recover $\yg$ from~\eqref{eq:PH_w_def}.

\emph{Particular solution.} Since $\bm{P}$ is banded block Toeplitz,
\eqref{eq:PH_PART} is solved by the forward substitution
\begin{equation}
\label{eq:particular_recurrence}
\wbp\ssub{m} = \vbp\ssub{m}
  - a\ssub{1}\,\wbp\ssub{m-1} - a\ssub{2}\,\wbp\ssub{m-2},
\quad m = 0, \ldots, N{-}1,
\end{equation}
with $\wbp\ssub{-1} = \wbp\ssub{-2} = \bm{0}$. Since each block
requires the two preceding blocks, the forward substitution has
sequential depth $\mathcal{O}(N)$. Note that neither $\vg$
nor~\eqref{eq:particular_recurrence} involves $\yb\ssub{-1}$: the
particular solution of every group can be computed independently.

\emph{Recovering the output.} The matrix $\bm{P}^{-1}\Agm$
in~\eqref{eq:PH_w_def} can be derived as the explicit form
\begin{equation}
\label{eq:P_inv_A_int}
\bm{P}^{-1}\Agm =
\begin{bmatrix}
u\ssub{2,0}\,\mathbf{e}\ssub{0} & u\ssub{1,0}\,\mathbf{e}\ssub{0} \\
u\ssub{2,1}\,\mathbf{e}\ssub{0} & u\ssub{1,1}\,\mathbf{e}\ssub{0} \\
\vdots & \vdots \\
u\ssub{2,N-1}\,\mathbf{e}\ssub{0} & u\ssub{1,N-1}\,\mathbf{e}\ssub{0}
\end{bmatrix},
\end{equation}
where $\mathbf{e}\ssub{0} = [1,\, 0,\, \ldots,\, 0]^\top \in
\mathbb{R}^{L}$ and the coefficients $u\ssub{l,m}$ follow the same
recurrence as those in $\bm{H}\ssub{12}$ of $\bm{H}$
\big[\!\cite{Zhai_26}, Eq.~(21)\big]. Combining~\eqref{eq:PH_w_def}
with~\eqref{eq:P_inv_A_int}, the first $N{-}2$ blocks satisfy
\begin{equation}
\label{eq:PH_forward}
\ybp\ssub{m}
+ u\ssub{2,m}\,\ybp\ssub{-2}
+ u\ssub{1,m}\,\ybp\ssub{-1}
= \wbp\ssub{m},
\quad m = 0, \ldots, N{-}3,
\end{equation}
where
$\ybp\ssub{-2} = [y[-2],\, y[N{-}2],\, \ldots,\,
y[(L{-}1)N{-}2]]^\top$ and
$\ybp\ssub{-1} = [y[-1],\, y[N{-}1],\, \ldots,\,
y[(L{-}1)N{-}1]]^\top$, i.e., the terminal blocks shifted by one
position with the initial state in the first entry. The last two
blocks satisfy
\begin{equation}
\label{eq:recursive_doubling_after_permute}
\begin{bmatrix}
\bm{I}  &          &          &        \\
\bm{C}  & \bm{I}   &          &        \\
        & \ddots   & \ddots   &        \\
        &          & \bm{C}   & \bm{I}
\end{bmatrix}
\begin{bmatrix}
\ye[0] \\ \ye[1] \\ \vdots \\ \ye[L{-}1]
\end{bmatrix}
+
\begin{bmatrix}
\bm{C} \\ \bm{0} \\ \vdots \\ \bm{0}
\end{bmatrix}
\yb\ssub{-1}
=
\begin{bmatrix}
\we[0] \\ \we[1] \\ \vdots \\ \we[L{-}1]
\end{bmatrix},
\end{equation}
where $\ye[l] = [y[(l{+}1)N{-}2],\, y[(l{+}1)N{-}1]]^\top$, $\we[l]$
is defined analogously, and the entries of the $2 \times 2$ matrix
$\bm{C}$ are precomputed from the filter coefficients \big[\!\cite{Zhai_26},
Eq.~(30)\big].

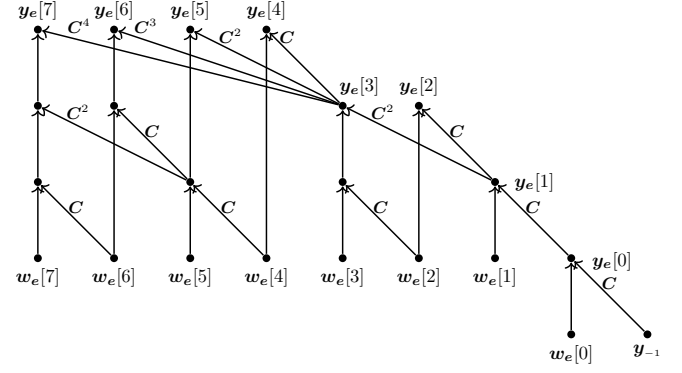
\begin{figure}[t]
\centering
\scalebox{0.7}{$
\begin{tikzpicture}[scale=0.72]

    \node[circle, fill=black, inner sep=1.5pt] (ye7) at (0, 6) {};
    \node[above=2pt] at (ye7) {$\ye[7]$};

    \node[circle, fill=black, inner sep=1.5pt] (ye6) at (2, 6) {};
    \node[above=2pt] at (ye6) {$\ye[6]$};

    \node[circle, fill=black, inner sep=1.5pt] (ye5) at (4, 6) {};
    \node[above=2pt] at (ye5) {$\ye[5]$};

    \node[circle, fill=black, inner sep=1.5pt] (ye4) at (6, 6) {};
    \node[above=2pt] at (ye4) {$\ye[4]$};

    \node[circle, fill=black, inner sep=1.5pt] (l2n1) at (0, 4) {};
    \node[circle, fill=black, inner sep=1.5pt] (l2n2) at (2, 4) {};

    \node[circle, fill=black, inner sep=1.5pt] (ye3) at (8, 4) {};
    \node[above=2pt, xshift=9pt] at (ye3) {$\ye[3]$};

    \node[circle, fill=black, inner sep=1.5pt] (ye2) at (10, 4) {};
    \node[above=2pt] at (ye2) {$\ye[2]$};

    \node[circle, fill=black, inner sep=1.5pt] (l3n1) at (0, 2) {};
    \node[circle, fill=black, inner sep=1.5pt] (l3n3) at (4, 2) {};
    \node[circle, fill=black, inner sep=1.5pt] (l3n5) at (8, 2) {};

    \node[circle, fill=black, inner sep=1.5pt] (ye1) at (12, 2) {};
    \node[right=2pt,xshift=5pt] at (ye1) {$\ye[1]$};

    \node[circle, fill=black, inner sep=1.5pt] (we7) at (0, 0) {};
    \node[below=2pt] at (we7) {$\we[7]$};

    \node[circle, fill=black, inner sep=1.5pt] (we6) at (2, 0) {};
    \node[below=2pt] at (we6) {$\we[6]$};

    \node[circle, fill=black, inner sep=1.5pt] (we5) at (4, 0) {};
    \node[below=2pt] at (we5) {$\we[5]$};

    \node[circle, fill=black, inner sep=1.5pt] (we4) at (6, 0) {};
    \node[below=2pt] at (we4) {$\we[4]$};

    \node[circle, fill=black, inner sep=1.5pt] (we3) at (8, 0) {};
    \node[below=2pt] at (we3) {$\we[3]$};

    \node[circle, fill=black, inner sep=1.5pt] (we2) at (10, 0) {};
    \node[below=2pt] at (we2) {$\we[2]$};

    \node[circle, fill=black, inner sep=1.5pt] (we1) at (12, 0) {};
    \node[below=2pt] at (we1) {$\we[1]$};

    \node[circle, fill=black, inner sep=1.5pt] (ye0) at (14, 0) {};
    \node[below=2pt, xshift=22pt, yshift=9pt] at (ye0) {$\ye[0]$};

    \node[circle, fill=black, inner sep=1.5pt] (init1) at (14, -2) {};

    \node[circle, fill=black, inner sep=1.5pt] (init0) at (16, -2) {};

    \draw[->, thick] (init0) -- (ye0) node[midway, above, font=\small] {$\bm{C}$};
    \node[below=2pt] at (init0) {$\yb\ssub{-1}$};

    \draw[->, thick] (init1) -- (ye0);
    \node[below=2pt] at (init1) {$\we[0]$};

    \draw[->, thick] (we7) -- (l3n1);
    \draw[->, thick] (we6) -- (l2n2);
    \draw[->, thick] (we5) -- (l3n3);
    \draw[->, thick] (we3) -- (l3n5);
    \draw[->, thick] (we1) -- (ye1);

    \draw[->, thick] (l3n1) -- (l2n1);
    \draw[->, thick] (l3n5) -- (ye3);
    \draw[->, thick] (we2) -- (ye2);

    \draw[->, thick] (l2n1) -- (ye7);
    \draw[->, thick] (l2n2) -- (ye6);
    \draw[->, thick] (l3n3) -- (ye5);
    \draw[->, thick] (we4) -- (ye4);

    \draw[->, thick] (we6) -- (l3n1) node[midway, above , font=\small] {$\bm{C}$};

    \draw[->, thick] (we4) -- (l3n3) node[midway, above , font=\small] {$\bm{C}$};

    \draw[->, thick] (we2) -- (l3n5) node[midway, above , font=\small] {$\bm{C}$};

    \draw[->, thick] (ye0) -- (ye1) node[midway, above, font=\small] {$\bm{C}$};

    \draw[->, thick] (l3n3) -- (l2n1) node[near end, above , font=\small] {$\bm{C}^2$};

    \draw[->, thick] (l3n3) -- (l2n2) node[midway, above , font=\small] {$\bm{C}$};

    \draw[->, thick] (ye1) -- (ye2) node[midway, above , font=\small] {$\bm{C}$};

    \draw[->, thick] (ye1) -- (ye3) node[near end, above , font=\small] {$\bm{C}^2$};

    \draw[->, thick] (ye3) -- (ye7) node[very near end, above , font=\small] {$\bm{C}^4$};

    \draw[->, thick] (ye3) -- (ye6) node[very near end, above , font=\small] {$\bm{C}^3$};

    \draw[->, thick] (ye3) -- (ye5) node[near end, above , font=\small] {$\bm{C}^2$};

    \draw[->, thick] (ye3) -- (ye4) node[near end, above , font=\small] {$\bm{C}$};

\end{tikzpicture}
$}
\caption{Sklansky construction~\cite{Sklansky_60} for the matrix
scan~\eqref{eq:rd_scan} with $L = 8$. The initial state
$\yb\ssub{-1}$ of the group enters the network at $\ye[0]$.}
\label{fig:recursive_doubling_diagram}
\end{figure}

\emph{Recursive doubling.} Each row
of~\eqref{eq:recursive_doubling_after_permute} is the first-order
matrix recurrence
\begin{equation}
\label{eq:rd_scan}
\ye[l] = \we[l] - \bm{C}\,\ye[l-1], \quad l = 0, \ldots, L{-}1,
\end{equation}
with $\ye[-1] = \yb\ssub{-1}$. The recurrence
is solved by recursive doubling with the Sklansky
construction~\cite{Sklansky_60}, which takes $\log_2 L$ levels.
Figure~\ref{fig:recursive_doubling_diagram} illustrates this
construction for $L = 8$, with the initial state $\yb\ssub{-1}$
entering the network at $\ye[0]$. The efficient computation of the
recursive doubling structure shown in
Figure~\ref{fig:recursive_doubling_diagram} is developed in the
Part~I~\cite{Zhai_26}.

\emph{Homogeneous correction.} 
With the terminal blocks and the
initial state ($\yb\ssub{-1}$) known, the remaining $N{-}2$ blocks are
recovered from~\eqref{eq:PH_forward} independently of signal groups. 

Note that in Part~I~\cite{Zhai_26}, since the focus is single group algorithm computation, we assume the initial state $\yb\ssub{-1}$ is known when a group starts.
Then, recursive doubling and the homogeneous solution~\eqref{eq:PH_forward} follow
the particular solution directly. Under parallel group execution,
however, every group other than the first obtains $\yb\ssub{-1}$
from the recursive doubling of its predecessor: later groups cannot
complete their homogeneous solution until the terminal states of all
earlier groups are known. The terminal states must therefore
propagate across the groups first; once this inter-group state
propagation completes, the homogeneous solutions of all groups are
independent and proceed fully in parallel. We will discuss the principle of inter-group state propagation later
in Section~\ref{sec:bg:prop}. 

\subsection{Cyclic Reduction}
\label{sec:bg:cr}

Cyclic reduction in \cite{Zhai_26} solves~\eqref{eq:recursive} by
eliminating alternating blocks. It proceeds in three steps:
reduction, terminal block filtering, and back substitution.

\emph{Reduction.} Each reduction level removes the even-indexed
blocks of the current system and produces a system of half the size
with the same second-order form. The elimination of the left-hand
side $\Ag$ is developed in Part~I~\cite{Zhai_26}; here, the right-hand side additionally carries the
initial-state term $\Agm\,\yb\ssub{-1}$ of~\eqref{eq:recursive}.
After the same sequence of permutations and eliminations, the system
at level $i$ becomes
\begin{equation}
\label{eq:cyclic_reduction_with_states_odd}
\Ag\ssub{CR}^{(i)}\,\yg^{(i)}
= \vg^{(i)}
- \Agm^{(i)}\,\yb\ssub{-1},
\end{equation}
where $\Agm^{(i)}$ is the initial-condition matrix at reduction
level $i$, given by
\begin{equation}
\label{eq:cyclic_reduction_A_minus}
\renewcommand{\arraystretch}{0.8}
\setlength\arraycolsep{2pt}
\Agm^{(i)} =
\left[
\begin{array}{cc}
h^{(i)}\mathbf{e\ssub{0}}  & g^{(i)}\mathbf{e\ssub{0}} \\
   & d^{(i)}\mathbf{e\ssub{0}} \\
   & \\
   & \\
\hdashline
 h^{(i+1)}\mathbf{e\ssub{0}}  & g^{(i+1)}\mathbf{e\ssub{0}} \\
   & f^{(i+1)}\mathbf{e\ssub{0}} \\
   & \\
   &
\end{array}
\right],
\end{equation}
and $d^{(i)}$ is the updated coupling coefficient of the
even-indexed blocks \big[\!\cite{Zhai_26}, Eq.~(41)\big]. The lower half of
$\Agm^{(i)}$ carries over to the next level as $\Agm^{(i+1)}$,
mirroring the reduction of $\Ag\ssub{CR}^{(i)}$ itself. This pattern
continues until the final level $i = \log\ssub{2} N - 1$, where
\begin{equation}
\label{eq:cyclic_reduction_recursive_odd_last_with_state}
\renewcommand{\arraystretch}{0.8}
\setlength\arraycolsep{2pt}
\Agm^{(i)} =
\begin{bmatrix}
 h^{(i)}\mathbf{e\ssub{0}}  & (g^{(i)}+d^{(i)}\SR)\mathbf{e\ssub{0}} \\[2pt]
 h^{(i+1)}\mathbf{e\ssub{0}}  & (g^{(i+1)}+f^{(i+1)}\SR)\mathbf{e\ssub{0}} \\
\end{bmatrix}.
\end{equation}
The initial-condition coefficients are updated alongside the reduced
coefficients $e^{(i)}$ and $f^{(i)}$, with $e^{(0)} = a\ssub{1}$ and
$f^{(0)} = a\ssub{2}$ \big[\!\cite{Zhai_26}, Eq.~(41)\big], as
\begin{equation}
\label{eq:cyclic_reduction_odd_coef_with_state}
\begin{aligned}
h^{(0)} &= a\ssub{2}, &\quad g^{(0)} &= a\ssub{1}, \\
h^{(i+1)} &= -e^{(i)}h^{(i)}, &\quad
g^{(i+1)} &= f^{(i)}-e^{(i)}g^{(i)}.
\end{aligned}
\end{equation}
All coefficients depend only on the filter parameters and are
precomputed. The reduction proceeds over $\log_2 N$ levels; all
block updates within a level are independent.

\emph{Terminal block filtering.} At the deepest level, a single
block remains: the coupling of $\ybp\ssub{N-1}$ to all other blocks
has been eliminated, and the lower row
of~\eqref{eq:cyclic_reduction_recursive_odd_last_with_state} yields
\begin{equation}
\label{eq:cr_terminal_block}
\scalebox{0.86}{$
\begin{aligned}
\bigl(\bm{I} {+} e^{(\log\ssub{2} N)}\,\SR
  {+} f^{(\log\ssub{2} N)}\,\SR^{2}\bigr)\,\ybp\ssub{N-1}
&{=} \vbp\ssub{N-1}^{(\log\ssub{2} N)} 
{-} h^{(\log\ssub{2} N)}\,\mathbf{e\ssub{0}}\,y[-2] \\
&{-} \bigl(g^{(\log\ssub{2} N)}{+}f^{(\log\ssub{2} N)}\,\SR\bigr)\,
  \mathbf{e\ssub{0}}\,y[-1],
\end{aligned}$}
\end{equation}
where $\vbp\ssub{N-1}^{(\log\ssub{2} N)}$ is the accumulated
right-hand side of the last block after all $\log_2 N$ reduction
levels. Since the left hand side
in~\eqref{eq:cr_terminal_block} is a unit banded lower-triangular
Toeplitz matrix, it can be solved by block filtering.

\emph{Back substitution.} With $\yb\ssub{-1} = [y[-2],\, y[-1]]^\top$ known, the back
substitution recovers the eliminated blocks level by level in
reverse order of the reduction. All recoveries within a level are
independent, and the even-indexed blocks of the original
system---including $\ybp\ssub{N-2}$---are recovered at the last
level.

Cyclic reduction requires approximately $6N$ block FMAs per
group, with sequential depth $\mathcal{O}(\log_2 N)$ in both the
reduction and the back substitution. Owing to this shorter
dependency chain compared to the $\mathcal{O}(N)$ particular
solution of the PH factorization~\eqref{eq:particular_recurrence},
the SIMD implementation of cyclic reduction achieved the best
throughput on a single CPU core shown in \cite{Zhai_26}.

\subsection{Divide-and-Conquer Cyclic Reduction}
\label{sec:bg:dccr}

The cyclic reduction in \cite{Zhai_26} is one-sided: each
level eliminates one parity and retains the other, so the eliminated
blocks are recovered only during back substitution. In particular,
the terminal state of the group is split across the two parities:
$\ybp\ssub{N-1}$ emerges at the deepest reduction level, but its
companion $\ybp\ssub{N-2}$ is recovered only at the \emph{last}
substitution level. Cyclic reduction is, at its core, a
divide-and-conquer method~\cite{Walter_97}, and the classical device
for extracting more parallelism from it is to apply the elimination
around both parities, trading redundant arithmetic for a second,
independent subproblem.

Unlike the preceding algorithms, the divide-and-conquer variant does
not evaluate the non-recursive stage~\eqref{eq:feedforward}
separately. Instead, the elimination is applied to both sides
of~\eqref{eq:multiblock_system}, reducing $\Ag$ and $\Bg$ together,
so that the non-recursive computation is folded into the reduction
steps. Grouping the blocks by parity and eliminating the
even-indexed blocks yields the odd-indexed subsystem
\begin{equation}
\label{eq:dc_odd}
\Ag\ssub{CR}^{(1)}\,\yg\ssub{\mathrm{odd}}
= \Bg\ssub{\mathrm{main}}^{(1)}\,\xg\ssub{\mathrm{odd}}
+ \Bg\ssub{\mathrm{cross}}^{(1)}\,\xg\ssub{\mathrm{even}},
\end{equation}
where $\yg\ssub{\mathrm{odd}} = [\ybp\ssub{1},\, \ybp\ssub{3},\,
\ldots,\, \ybp\ssub{N-1}]^\top$ and $\xg\ssub{\mathrm{odd}}$,
$\xg\ssub{\mathrm{even}}$ collect the odd- and even-indexed input
blocks. The matrix $\Ag\ssub{CR}^{(1)}$ is the level-1 reduced
system of \big[\!\cite{Zhai_26},
Eq.~(40)\big]. On the right-hand side, the main
matrix $\Bg\ssub{\mathrm{main}}^{(1)}$ carries the contribution of
the retained input blocks and has the same block structure as
$\Ag\ssub{CR}^{(1)}$, with coefficients $e\ssub{B}^{(1)}$ and
$f\ssub{B}^{(1)}$; the cross matrix $\Bg\ssub{\mathrm{cross}}^{(1)}$
carries the contribution of the eliminated input blocks, with
coefficients $c\ssub{B}^{(1)}$ and $d\ssub{B}^{(1)}$:
\begin{equation}
\label{eq:dc_B_matrices}
\renewcommand{\arraystretch}{0.85}
\setlength\arraycolsep{1.5pt}
\begin{aligned}
\Bg\ssub{\mathrm{main}}^{(1)}
&=
\left[
\begin{array}{cccc}
  \bm{I}  &   & f\ssub{B}^{(1)}\SR  & e\ssub{B}^{(1)}\SR  \\[2pt]
 e\ssub{B}^{(1)}\bm{I}  & \bm{I}  &   & f\ssub{B}^{(1)}\SR  \\[2pt]
 f\ssub{B}^{(1)}\bm{I}  & \ddots  & \ddots  &   \\[2pt]
  & \ddots  & e\ssub{B}^{(1)}\bm{I}  & \bm{I}
\end{array}
\right], \\[4pt]
\Bg\ssub{\mathrm{cross}}^{(1)}
&=
\left[
\begin{array}{cccc}
 d\ssub{B}^{(1)}\bm{I}  &   &   & c\ssub{B}^{(1)}\SR  \\[2pt]
 c\ssub{B}^{(1)}\bm{I}  & d\ssub{B}^{(1)}\bm{I}  &   &   \\[2pt]
  & \ddots  & \ddots  &   \\[2pt]
  &   & c\ssub{B}^{(1)}\bm{I}  & d\ssub{B}^{(1)}\bm{I}
\end{array}
\right].
\end{aligned}
\end{equation}
Symmetrically, eliminating the odd-indexed blocks yields the
even-indexed subsystem
\begin{equation}
\label{eq:dc_even}
\Ag\ssub{CR}^{(1)}\,\yg\ssub{\mathrm{even}}
= \Bg\ssub{\mathrm{main}}^{(1)}\,\xg\ssub{\mathrm{even}}
+ \widetilde{\Bg}\ssub{\mathrm{cross}}^{(1)}\,\xg\ssub{\mathrm{odd}},
\end{equation}
with $\yg\ssub{\mathrm{even}} = [\ybp\ssub{0},\, \ybp\ssub{2},\,
\ldots,\, \ybp\ssub{N-2}]^\top$. The left-hand-side matrix and the
main matrix are identical to those in~\eqref{eq:dc_odd}, since both
reductions apply the same Schur complement to $\Ag$ and $\Bg$. The
cross matrix uses the same coefficients $c\ssub{B}^{(1)}$ and
$d\ssub{B}^{(1)}$, with the banding shifted by the reversed
elimination direction,
\begin{equation}
\label{eq:dc_B_cross_even}
\renewcommand{\arraystretch}{0.85}
\setlength\arraycolsep{1.5pt}
\widetilde{\Bg}\ssub{\mathrm{cross}}^{(1)}
=
\left[
\begin{array}{cccc}
     &   & c\ssub{B}^{(1)}\SR  & d\ssub{B}^{(1)}\SR  \\[2pt]
 d\ssub{B}^{(1)}\bm{I}  &   &   & c\ssub{B}^{(1)}\SR  \\[2pt]
 c\ssub{B}^{(1)}\bm{I}  & \ddots  &   &   \\[2pt]
  & \ddots  & d\ssub{B}^{(1)}\bm{I}  &
\end{array}
\right].
\end{equation}
The feedforward coefficients after the first reduction level are
\begin{equation}
\label{eq:dc_coef_B}
\begin{aligned}
f\ssub{B}^{(0)} &= b\ssub{2} &\quad
e\ssub{B}^{(0)} &= b\ssub{1} \\
f\ssub{B}^{(1)} &= f^{(0)}\,f\ssub{B}^{(0)} &\quad
e\ssub{B}^{(1)} &= f\ssub{B}^{(0)} + f^{(0)}
- e^{(0)}\,e\ssub{B}^{(0)} \\
c\ssub{B}^{(1)} &= f^{(0)}\,e\ssub{B}^{(0)}
- e^{(0)}\,f\ssub{B}^{(0)} &\quad
d\ssub{B}^{(1)} &= e\ssub{B}^{(0)} - e^{(0)}
\end{aligned}
\end{equation}
where $e^{(0)} = a\ssub{1}$ and $f^{(0)} = a\ssub{2}$. The
initial-condition and wrap-around terms reduce alongside, through
initial-condition matrices of the same form
as~\eqref{eq:cyclic_reduction_A_minus}, and are omitted here for
clarity.

In this work, one level of divide-and-conquer is performed. The two
subsystems~\eqref{eq:dc_odd} and~\eqref{eq:dc_even} are independent,
and each is solved by the one-sided cyclic reduction on $N/2$ blocks, starting from level $i =
1$. When two processors are available, the two subsystems proceed
concurrently, which is the first benefit of the divide-and-conquer
form: it doubles the parallelism of every reduction and substitution
level.

The second benefit concerns the terminal state. The odd subsystem
exposes $\ybp\ssub{N-1}$ at its deepest reduction level, and the
even subsystem exposes $\ybp\ssub{N-2}$. Both terminal blocks---and
with them the terminal state $[y[NL{-}2],\, y[NL{-}1]]^\top$ of the
group---are therefore available before any back substitution begins.
This property is essential for parallel group execution: the
terminal states can first be propagated across all groups, after
which the back substitution of every group depends only on data
within the group, and all groups proceed independently in parallel.

The added parallelism is paid for in arithmetic: reducing both
parities requires approximately $4N$ block FMAs per direction and
$8N$ in total, against $6N$ for the one-sided variant, with the
block shuffle count roughly doubled as well. Divide-and-conquer
cyclic reduction is therefore worthwhile exactly when parallel
resources can absorb the redundancy, as in the batched GPU execution
of Section~\ref{sec:gpu}.

\subsection{Inter-Group State Propagation}
\label{sec:bg:prop}

The algorithms above compute a single signal block group, and the
initial state $\yb\ssub{-1}$ of each group is the terminal output of
its predecessor. The straightforward execution is therefore
sequential: the groups are processed in order, and the overall rate
is set by the per-group throughput of the chosen algorithm, which is
the basis of comparison in Part~I~\cite{Zhai_26}. Parallel
execution across groups requires breaking this chain: the initial
conditions of all groups must be resolved before, or independently
of, the bulk of the per-group computation. We refer to this as
inter-group state propagation.

The scan~\eqref{eq:rd_scan} includes the initial state $\ye[-1] =
\yb\ssub{-1}$, which depends on the terminal output of the preceding
group. Two strategies exist for incorporating this term, shown in
Figure~\ref{fig:recursive_doubling_diagram}. The first
includes $\yb\ssub{-1}$ at the initial level, so that the recursion
propagates its contribution through all subsequent levels. The
second sets $\ye[-1] = \bm{0}$ and runs the recursive doubling
independently of the initial state, producing the zero-state
response. Since~\eqref{eq:rd_scan} is linear, the contribution of
$\yb\ssub{-1}$ can be superposed afterward as
\begin{equation}
\label{eq:RD_initial_state_correction}
\ye[l] = \bm{y_{e,0}}[l] - \bm{C}^{l+1}\,\yb\ssub{-1},
\quad l = 0, 1, \ldots, L{-}1,
\end{equation}
where $\bm{y_{e,0}}[l]$ denotes the zero-state output. Both
strategies have the same computational complexity, but the second is
preferable for parallel architectures: the zero-state recursive
doubling requires no inter-group information and can be executed
independently across all groups, with the
correction~\eqref{eq:RD_initial_state_correction} applied once the
initial states are resolved. The same decomposition applies to
cyclic reduction, where the initial state enters the terminal
solve~\eqref{eq:cr_terminal_block} only through the explicit
precomputed terms.

The structure of Figure~\ref{fig:recursive_doubling_diagram} makes
the resolution of the initial states themselves equally simple.
Setting $l = L{-}1$ in~\eqref{eq:RD_initial_state_correction}, the
terminal state of a group is its zero-state terminal vector
corrected by $-\bm{C}^{L}\,\yb\ssub{-1}$. To track this across
groups, index the groups by $g$ and append the index as a second
argument: $\bm{y_{e,0}}[l,\, g]$ denotes the zero-state output of
group $g$, and $\yb\ssub{-1}[g]$ its initial state. Writing
$\bm{t}[g] = \bm{y_{e,0}}[L{-}1,\, g]$ for the zero-state terminal
vector of group $g$, the initial states of consecutive groups
therefore satisfy
\begin{equation}
\label{eq:inter_group_recurrence}
\yb\ssub{-1}[g{+}1] = \bm{t}[g] - \bm{C}^{L}\,\yb\ssub{-1}[g],
\end{equation}
which is a first-order matrix recurrence of exactly the same form
as~\eqref{eq:rd_scan}, one level up: the groups themselves form an
upper-level recursive doubling, with $\bm{C}^{L}$ in place of
$\bm{C}$ and the zero-state terminal vectors in place of $\we[l]$.

The upper level admits the same zero-state treatment in turn.
Unrolling~\eqref{eq:inter_group_recurrence} expresses each
$\yb\ssub{-1}[g]$ as a combination of the zero-state terminal
vectors $\bm{t}[0], \ldots, \bm{t}[g{-}1]$ and the initial state
$\yb\ssub{-1}[0]$ of the first group alone. Consequently, once the
zero-state terminal vectors of all groups and the initial condition
of the first group are known, the true initial condition of every
group can be computed independently---sequentially, or by one more
recursive doubling across the groups.

The zero-state terminal vector of a group is commonly referred to as
its \emph{local carry}, and the resolved initial condition as its
\emph{global carry}: local carries are computed fully in parallel,
and global carries are resolved by a lightweight propagation whose
cost is independent of the group size. The same strategy underlies
the GPU implementation discussed in Section~\ref{sec:gpu}.

\section{Real-Time Processing on Multi-Core CPUs}
\label{sec:cpu}

\noindent The single-core benchmarks in Part~I~\cite{Zhai_26}
established the best achievable throughput for each computational
kernel. This section composes those optimized kernels into a
multi-core implementation targeting real-time signal processing,
where input data must be processed in FIFO order with minimal
latency. We first present the wavefront execution strategy, then its
realization with TBB flow graphs, and finally the multi-core
measurements on Meteor Lake.

\subsection{Wavefront Execution}
\label{sec:wavefront}

\begin{figure}[t]
\centering
\scalebox{0.95}{
\begin{tikzpicture}[
    node distance=0.75cm and 0.75cm,
    biquad/.style={circle, draw, minimum size=0.45cm, inner sep=0pt, font=\tiny},
    arr/.style={->, >=stealth, thin}
]
\definecolor{t1}{RGB}{228,26,28}
\definecolor{t2}{RGB}{55,126,184}
\definecolor{t3}{RGB}{77,175,74}
\definecolor{t4}{RGB}{152,78,163}
\definecolor{t5}{RGB}{255,127,0}
\foreach \g [count=\gi from 0] in {0,...,4} {
  \foreach \s [count=\si from 0] in {0,...,4} {
    \pgfmathtruncatemacro{\tidx}{mod(\si,5)}
    \ifcase\tidx
      \def\tcolor{t1}\or
      \def\tcolor{t2}\or
      \def\tcolor{t3}\or
      \def\tcolor{t4}\or
      \def\tcolor{t5}
    \fi
    \pgfmathtruncatemacro{\diag}{\gi+\si}
    \node[biquad, fill=\tcolor!40] (n\gi\si) at (\si*0.9, -\diag*0.9) {};
  }
}
\foreach \g in {0,...,4} {
  \foreach \s [evaluate=\s as \snext using int(\s+1)] in {0,...,3} {
    \draw[arr, gray!50] (n\g\s) -- (n\g\snext);
  }
}
\foreach \s in {0,...,4} {
  \foreach \g [evaluate=\g as \gnext using int(\g+1)] in {0,...,3} {
    \draw[arr] (n\g\s) -- (n\gnext\s);
  }
}
\node[above=0.1cm of n00, font=\scriptsize] {group 0};
\node[above=0.1cm of n01, font=\scriptsize] {group 1};
\node[above=0.1cm of n04, font=\scriptsize] {group 4};
\node[left=0.1cm of n00, font=\scriptsize] {$F_1$};
\node[left=0.1cm of n10, font=\scriptsize] {$F_2$};
\node[left=0.1cm of n40, font=\scriptsize] {$F_K$};
\draw[arr, thick] (-1.2,-0.2) -- (-1.2,-4.5) node[midway, left, font=\scriptsize, rotate=90, anchor=south] {filter order / time};
\draw[arr, thick] (-0.4,1.0) -- (3.8,1.0) node[midway, above, font=\scriptsize] {signal group / core};
\end{tikzpicture}
}
\caption{Wavefront execution of a cascaded IIR filter with $K$
second-order sections. Each circle is one section applied to one
signal block group; colors indicate thread assignment, with one
thread following each group through the entire cascade. The signal
flows vertically through the cascade, section states pass diagonally
to the next group, and nodes at the same height execute
concurrently.}
\label{fig:wavefront}
\end{figure}
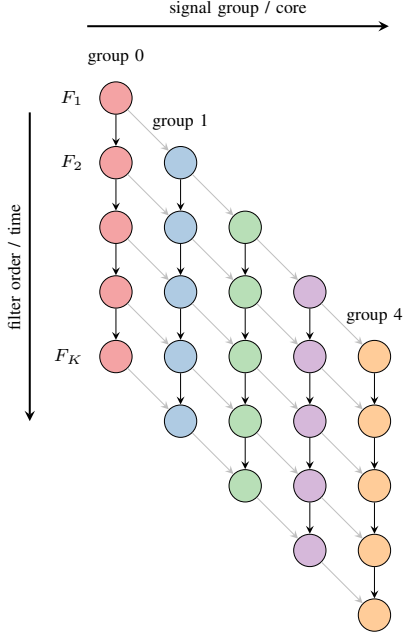

We adopt a wavefront execution strategy for multi-core CPU
implementations of cascaded IIR filters.
Figure~\ref{fig:wavefront} illustrates the execution pattern for a
cascade of $K$ second-order sections. Each circle represents one
section applied to one signal block group; each column is a group,
and colors indicate thread assignment. Within a group, the signal
flows vertically down the cascade; between adjacent groups, each
section passes its output state along the gray diagonal to the next
group. Time advances downward, so all nodes at the same height
execute concurrently---one wavefront.

Wavefront execution possesses several properties that make it well
suited for real-time processing on multi-core CPUs. First, the
pipelined cascade structure preserves FIFO ordering by construction:
each signal block group enters and exits the cascade in the same
order. Second, the same signal block group is processed by a single
thread across all second-order sections, maximizing data locality
and minimizing inter-core context switching. Third, within each
section the groups execute strictly in order, so the initial state
is always available when a group arrives: the serial single-group
functions of \cite{Zhai_26} apply unchanged, and no inter-group state
propagation is required. A downstream section waits only for the
state of its predecessor, not for the entire cascade, so the
pipeline fill depth is proportional to the number of sections and
the steady-state throughput scales with the number of available
cores.

\subsection{Implementation with TBB Flow Graphs}
\label{sec:tbb}

\begin{figure}[t]
\centering
\scalebox{1.0}{
\begin{tikzpicture}[
    node distance=0.55cm,
    block/.style={draw, rounded corners, minimum height=0.55cm, minimum width=0.8cm, font=\scriptsize, inner sep=2pt},
    arr/.style={->, >=stealth, thick}
]
\node[block] (src) {$f()$};
\node[block, right=of src] (pm) {$f(\xg)$};
\node[block, right=of pm] (i0) {$f(\xg)$};
\node[right=0.3cm of i0, font=\scriptsize] (dots) {$\cdots$};
\node[block, right=0.3cm of dots] (i7) {$f(\xg)$};
\node[block, right=of i7] (dp) {$f(\xg)$};
\node[block, right=of dp] (snk) {$f()$};

\draw[arr] (src) -- (pm);
\draw[arr] (pm) -- (i0);
\draw[arr] (i0) -- (dots);
\draw[arr] (dots) -- (i7);
\draw[arr] (i7) -- (dp);
\draw[arr] (dp) -- (snk);

\node[below=0.08cm of src, font=\scriptsize] {source};
\node[below=0.08cm of pm, font=\scriptsize] {permute};
\node[below=0.08cm of i0, font=\scriptsize] {iir$_0$};
\node[below=0.08cm of i7, font=\scriptsize] {iir$_7$};
\node[below=0.08cm of dp, font=\scriptsize] {deperm.};
\node[below=0.08cm of snk, font=\scriptsize] {sink};
\end{tikzpicture}
}
\caption{TBB flow graph for a 16th-order cascaded IIR filter.
Each node has serial concurrency to enable wavefront pipelining.}
\label{fig:tbb_graph}
\end{figure}
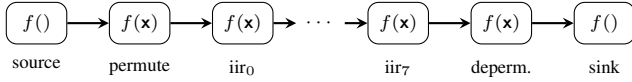

The wavefront execution is implemented with the Threading Building
Blocks (TBB)~\cite{Michael_19} flow graph, a task-based runtime for
expressing pipelined parallelism without explicit thread management.
The flow graph is a linear chain of second-order filter nodes, as
shown in Figure~\ref{fig:tbb_graph}, each configured with
\emph{serial concurrency}: only one thread can execute a given node
at any time, while different nodes execute concurrently across
threads, realizing the pipeline of Figure~\ref{fig:wavefront}. The
backbone of TBB is its task-queue-based scheduler: each worker
thread maintains a local task queue, allowing it to carry the same
signal block group through consecutive cascade stages whenever the
next stage is immediately available. This behavior aligns naturally
with the wavefront pattern.

\subsection{Function Size and Grain Size}
\label{sec:grain_size}

\begin{figure}[t]
    \centering
    \includegraphics[width=0.48\textwidth]{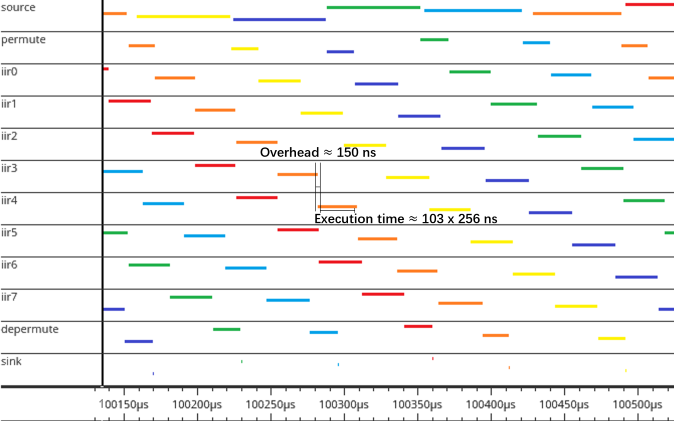}
    \caption{Node-level execution timeline of a 16th-order IIR filter
    on 6 Meteor Lake P-core threads, captured from TBB Flow Graph
    Analyzer. Each color indicates a different thread. The
    steady-state pipelining pattern and the approximately
    \SI{150}{\nano\second} inter-node overhead are visible.}
    \label{fig:tbb_pcore}
\end{figure}

Figure~\ref{fig:tbb_pcore} shows the steady-state execution timeline
of a 16th-order filter using the CR algorithm under the best
single-core configuration ($L = 8$, $N = 64$) on six P-core threads,
captured with the TBB Flow Graph Analyzer. A clear pipelining
pattern emerges: each thread consistently processes the same signal
block group, and approximately four filter nodes execute
concurrently across threads.

A key observation from Figure~\ref{fig:tbb_pcore} is that the
execution gap between adjacent nodes of the same thread---the
task-generation overhead of the runtime---is approximately
\SI{150}{\nano\second}. For CR-based multi-block filtering with a
function size of $NL = 512$ samples, the per-section latency on a
Meteor Lake P-core at \SI{4.5}{\giga\hertz} is
\[
\frac{7.3~\text{cycles/block}}{8~\text{samples/block}} \cdot
\frac{512~\text{samples}}{4.5 \times 10^{9}~\text{cycles/s}}
\approx \SI{103}{\nano\second},
\]
where 7.3~cycles per block of $L = 8$ samples is the per-section
single-core measurement of Part~I~\cite{Zhai_26}. Since this execution
time is comparable to the \SI{150}{\nano\second} task-generation
overhead, the optimized single-core function size is too small for
efficient per-node execution in the task system.

To address this, we introduce an additional parameter, the
\emph{grain size}---in TBB terminology~\cite{Michael_19}---which
specifies how many signal block groups are aggregated and processed
within a single node invocation. A grain size of 256 increases the
per-node execution time to $103 \times 256 \approx
\SI[parse-numbers=false]{26{,}400}{\nano\second} \gg
\SI{150}{\nano\second}$, effectively amortizing the task-generation
overhead. The grain size thus controls task granularity, at the cost
of a proportional increase in per-group latency. Function size and
grain size together form a two-level strategy: the function size
determines the efficiency of processing each signal block group on a
single core, while the grain size determines the task granularity
needed to hide the overhead of the runtime scheduler.

\subsection{Heterogeneous Cores}
\label{sec:ecore}

\begin{figure}[t]
    \centering
    \includegraphics[width=0.48\textwidth]{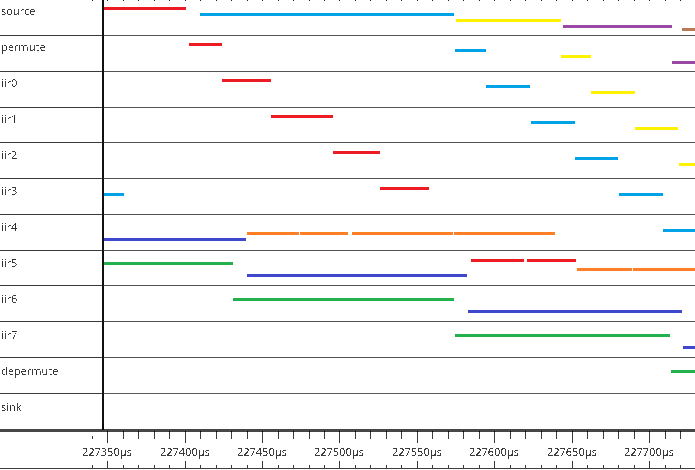}
    \caption{Execution timeline with 8 threads (6~P-cores +
    2~E-cores). The slower E-core nodes (longer segments) become the
    pipeline bottleneck.}
    \label{fig:tbb_mixed}
\end{figure}

Meteor Lake-H features a hybrid architecture with six hyperthreaded
performance cores (P-cores) and eight efficiency cores
(E-cores)~\cite{Lam_24}. E-cores support only 128-bit vector
registers, so an $L = 8$ single-precision block requires two vector
instructions instead of one on a P-core; they also run at a lower
clock, \SI{3.8}{\giga\hertz} against the P-cores'
\SI{4.5}{\giga\hertz}. Since the throughput of a
pipeline is determined by its slowest stage, E-core nodes become the
bottleneck. Figure~\ref{fig:tbb_mixed} shows the execution timeline
when two E-cores are introduced alongside six P-cores: the slower
E-core execution, visible as longer segments for
iir$_4$--iir$_7$, creates pipeline bubbles that reduce aggregate
throughput below the six-P-core configuration.

\subsection{Measurement and Results}
\label{sec:cpu_results}

The multi-core experiments target the Meteor Lake-H platform of
Part~I~\cite{Zhai_26}: six hyperthreaded P-cores at
\SI{4.5}{\giga\hertz} and eight E-cores at
\SI{3.8}{\giga\hertz}~\cite{Lam_24}. Each P-core has \SI{2}{\mega\byte}
of private L2 cache; each cluster of four E-cores shares
\SI{2}{\mega\byte}. The single-core optimized kernels from Part~I,
implemented in C++ with the Vector Class Library~\cite{Agner_04},
are composed into the TBB flow graph of Figure \ref{fig:tbb_graph} with
the CR-based multi-block filtering algorithm ($L = 8$, $N = 64$).
Threads are pinned to P-cores first; E-cores are used only when the
thread count exceeds six. Throughput is measured by processing a
sufficiently large number of signal block groups to reach steady
state, and is reported normalized to the single-core baseline of
Part~I.

\begin{figure}[t]
    \centering
    \includegraphics[width=0.48\textwidth]{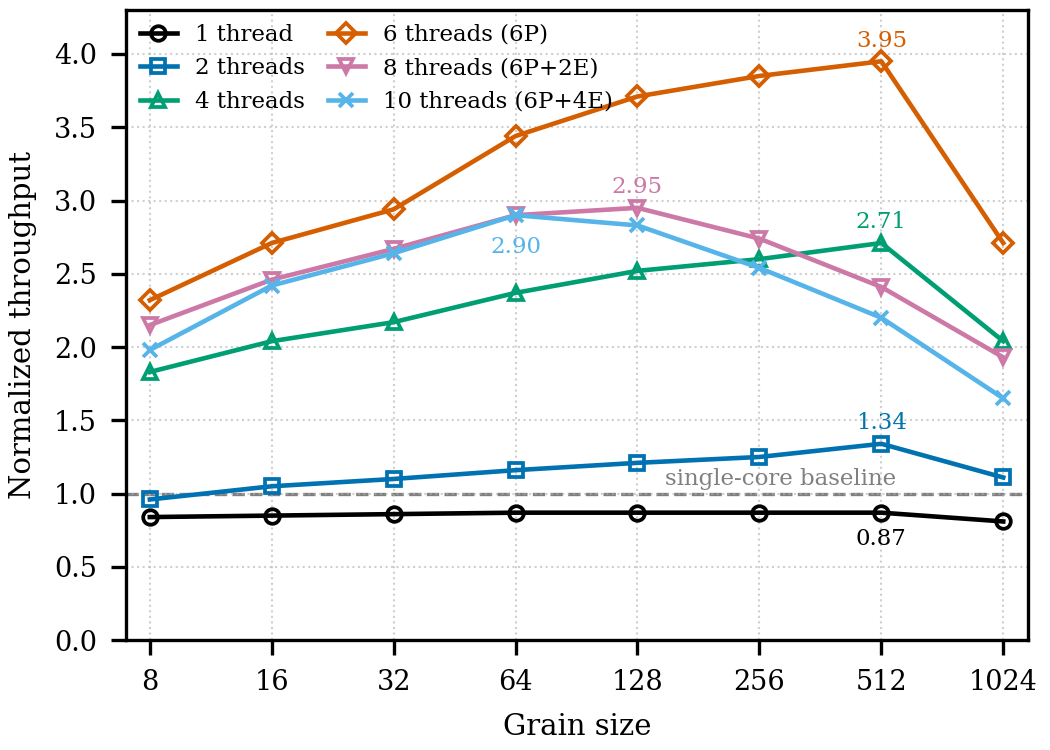}
    \caption{Normalized multi-core throughput of the 16th-order
    CR-based IIR filter on Meteor Lake versus grain size, for thread
    counts from 1 to 10. Throughput is the ratio to the single-core
    throughput (\SI{616}{\mega\samples\per\second}) of Part~I.}
    \label{fig:multicore_throughput}
\end{figure}

Figure~\ref{fig:multicore_throughput} presents the normalized
throughput as a function of grain size for one to ten threads, with
the largest value of each configuration marked next to its curve.
Four trends emerge. First, the single-thread curve approaches but
does not reach the baseline, saturating at 0.87; the remaining gap
is the TBB runtime overhead. Second, throughput increases with grain
size up to 512 as the task-generation overhead is amortized, then
drops sharply at 1024: the working set of 1024 groups of $NL = 512$
single-precision samples occupies $1024 \times 512 \times
4\,\mathrm{B} = \SI{2}{\mega\byte}$, reaching the private L2
capacity. Third, throughput scales with the thread count
and saturates at six P-cores, where the implementation reaches
$3.95\times$ at grain size 512. Fourth, adding E-cores degrades
performance: the eight-thread configuration (one E-core per cluster)
and the ten-thread configuration (two per cluster) fall below the
six-P-core curve at every grain size, as the slower E-core stages
bottleneck the pipeline (Figure~\ref{fig:tbb_mixed}). The
ten-thread configuration degrades at smaller grain sizes than the
eight-thread one because pairs of E-cores sharing the cluster L2
reach the effective per-core cache capacity earlier.
 
In absolute terms, the six-P-core configuration at grain size 512
achieves $3.95 \times \SI{616}{\mega\samples\per\second} \approx \SI{2.4}{\giga\samples\per\second}$
for the 16th-order system, where \SI{616}{\mega\samples\per\second}
is the single-core CR throughput of Part~I---itself an $8\times$
improvement over
\texttt{scipy.signal.sosfilt}~\cite{virtanen_20}.


\section{Batched Processing on GPUs}
\label{sec:gpu}

\noindent The multi-core implementation of Section~\ref{sec:cpu}
achieves low-latency streaming by pipelining signal block groups
across a small number of powerful cores. A GPU offers the opposite
resource: thousands of simple execution lanes whose collective
throughput is large, but only for work that decomposes into many
independent pieces. This section maps the multi-block filtering
algorithms onto that architecture and reports the measured results.

An NVIDIA GPU is an array of streaming multiprocessors (SMs).
Threads are scheduled in \emph{warps} of 32 that execute in lockstep;
warps are grouped into \emph{thread blocks} (TBs) that share a fast
on-chip memory (shared memory) and synchronize through barriers; TBs
are launched as a grid and cannot synchronize with each other except
through global (DRAM-backed) memory. When a warp stalls, the SM
switches to another resident warp at zero cost: latency is not
avoided but hidden by oversubscription~\cite{Volkov_16}. Arithmetic
is abundant; what costs is communication, at three price
levels---register shuffles inside a warp, barrier-synchronized shared
memory across warps, and global-memory round trips across TBs. The
experiments use the GTX~1070 (Pascal) and the RTX~3060 (Ampere);
Table~\ref{tab:devices} collects the hardware parameters used
throughout.

\begin{table}[!t]
\caption{The two test devices. The derived rows follow from
\eqref{eq:tmem} and \eqref{eq:wmin}.}
\label{tab:devices}
\centering
\footnotesize
\setlength{\tabcolsep}{4pt}
\begin{tabular}{@{}lcc@{}}
\toprule
 & GTX 1070 & RTX 3060 \\
\midrule
Architecture & Pascal GP104 & Ampere GA106 \\
SMs & 15 & 28 \\
Threads / warps / TBs per SM & 2048 / 64 / 32 & 1536 / 48 / 16 \\
Shared memory per SM / per TB & 96 / 48\,kB & 100 / 48\,kB \\
Registers per SM / per thread & 65536 / 255 & 65536 / 255 \\
FP32 cores per SM / FMA latency & 128 / 6\,cyc & 128 / 4\,cyc \\
Memory bandwidth & 256\,GB/s & 360\,GB/s \\
\midrule
Throughput roof \eqref{eq:tmem} & 32\,GS/s & 45\,GS/s \\
Hiding floor $W_{\min}$ \eqref{eq:wmin} & 8 warps & 6 warps \\
\bottomrule
\end{tabular}
\end{table}

\subsection{Thread-Block Mapping and Shared-Memory Staging}
\label{sec:gpu:mapping}

The input stream is divided into groups of $C = NL$ samples, and each
group is assigned to one thread block. Within its TB, the group is
carried through all $K$ sections of an order-$2K$ filter: the output of section
$s$, resident in registers, is the input of section $s{+}1$, so the
group is read from global memory once, filtered with $K$ second-order sections on chip,
and written once. This single-kernel, register-resident cascade
contrasts with prior GPU work on multi-block recursive
filtering~\cite{Lee_13}, which launches separate kernels per
stage---at least $2K{+}1$ launches for a $K$-section cascade---with
every intermediate result passing through global memory. Each TB
learns which group it owns by drawing a ticket from a global atomic
counter at launch; tickets ascend, so a TB's predecessors are always
scheduled no later than itself,  a property the state protocol of
inter-block lookback (Section~\ref{sec:gpu:lookback}) relies on.

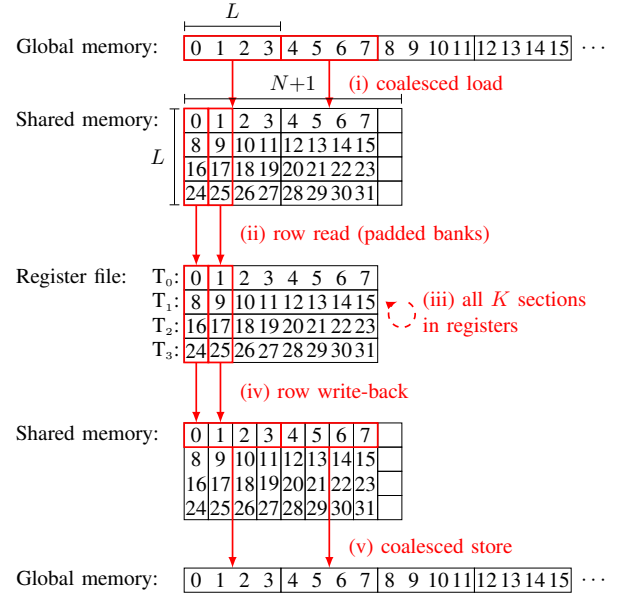
\begin{figure}[!t]
\centering
\scalebox{0.8}{  
\begin{tikzpicture}[scale=0.2]

\draw (0,0) rectangle (8,2);

\draw[|-|] (0,+2.8) -- (8,+2.8);
\node[above=0.01cm] at (4,+2.8) {$L$};

\node at (1,1) {0};
\node at (3,1) {1};
\node at (5,1) {2};
\node at (7,1) {3};

\begin{scope}[xshift=8cm]
    \draw (0,0) rectangle (8,2);
    \node at (1,1) {4};
    \node at (3,1) {5};
    \node at (5,1) {6};
    \node at (7,1) {7};
\end{scope}

\begin{scope}[xshift=16cm]
    \draw (0,0) rectangle (8,2);
    \node at (1,1) {8};
    \node at (3,1) {9};
    \node at (5,1) {10};
    \node at (7,1) {11};
\end{scope}

\begin{scope}[xshift=24cm]
    \draw (0,0) rectangle (8,2);
    \node at (1,1) {12};
    \node at (3,1) {13};
    \node at (5,1) {14};
    \node at (7,1) {15};
\end{scope}

\node[align=left] at (-8,1) {Global memory:};

\node at (34,1) {$\cdots$};

\draw (0,0)[red,thick] rectangle (8,2);

\node at (1,1) {};
\node at (3,1) {};
\node at (5,1) {};
\node at (7,1) {};

\begin{scope}[xshift=8cm]
    \draw[red,thick] (0,0) rectangle (8,2);
    \node at (1,1) {};
    \node at (3,1) {};
    \node at (5,1) {};
    \node at (7,1) {};
\end{scope}

\draw[-latex, red, thick] (4,0) -- (4,-4);
\draw[-latex, red, thick] (12,0) -- (12,-4);
\node[right, red] at (13.0,-2) {(i) coalesced load};

\begin{scope}[yshift=-6cm]

\begin{scope}[xshift=0cm]
    \draw (0,0) rectangle (8,2);
    \node at (1,1) {0};
    \node at (3,1) {1};
    \node at (5,1) {2};
    \node at (7,1) {3};

    \draw[|-|] (-0.8,2) -- (-0.8,-6);      
    \node at (-2.2,-2) {$L$};   
    \draw[|-|] (0,2.8) -- (18,2.8);
    \node[above=0.01cm] at (9,2.7) {$N{+}1$};
\end{scope}

\begin{scope}[xshift=8cm]
    \draw (0,0) rectangle (8,2);
    \node at (1,1) {4};
    \node at (3,1) {5};
    \node at (5,1) {6};
    \node at (7,1) {7};
    \draw (8,0) rectangle (10,2);
\end{scope}

\begin{scope}[xshift=0cm,yshift=-2cm]
    \draw (0,0) rectangle (8,2);
    \node at (1,1) {8};
    \node at (3,1) {9};
    \node at (5,1) {10};
    \node at (7,1) {11};
\end{scope}

\begin{scope}[xshift=8cm,yshift=-2cm]
    \draw (0,0) rectangle (8,2);
    \node at (1,1) {12};
    \node at (3,1) {13};
    \node at (5,1) {14};
    \node at (7,1) {15};
    \draw (8,0) rectangle (10,2);
\end{scope}

\begin{scope}[xshift=0cm,yshift=-4cm]
    \draw (0,0) rectangle (8,2);
    \node at (1,1) {16};
    \node at (3,1) {17};
    \node at (5,1) {18};
    \node at (7,1) {19};
\end{scope}

\begin{scope}[xshift=8cm,yshift=-4cm]
    \draw (0,0) rectangle (8,2);
    \node at (1,1) {20};
    \node at (3,1) {21};
    \node at (5,1) {22};
    \node at (7,1) {23};
    \draw (8,0) rectangle (10,2);
\end{scope}

\begin{scope}[xshift=0cm,yshift=-6cm]
    \draw (0,0) rectangle (8,2);
    \node at (1,1) {24};
    \node at (3,1) {25};
    \node at (5,1) {26};
    \node at (7,1) {27};
\end{scope}

\begin{scope}[xshift=8cm,yshift=-6cm]
    \draw (0,0) rectangle (8,2);
    \node at (1,1) {28};
    \node at (3,1) {29};
    \node at (5,1) {30};
    \node at (7,1) {31};
    \draw (8,0) rectangle (10,2);
\end{scope}

\node[align=left] at (-8,1) {Shared memory:};

\begin{scope}[xshift=0cm]
    \draw[red,thick] (0,-6) rectangle (2,2);
    \node at (1,1) {};
    \node at (1,-1) {};
    \node at (1,-3) {};
    \node at (1,-5) {};
\end{scope}

\begin{scope}[xshift=2cm]
    \draw[red,thick] (0,-6) rectangle (2,2);
    \node at (1,1) {};
    \node at (1,-1) {};
    \node at (1,-3) {};
    \node at (1,-5) {};
\end{scope}

\draw[-latex, red, thick] (1,-6) -- (1,-11);
\draw[-latex, red, thick] (3,-6) -- (3,-11);
\node[right, red] at (4.0,-8.5) {(ii) row read (padded banks)};

\end{scope}

\begin{scope}[yshift=-19cm]

\coordinate (sm_left) at (-11,2);

\coordinate (sm_right) at (34,2);    

\begin{scope}[xshift=0cm]
    \draw (0,0) rectangle (16,2);
    \node at (1,1) {0};
    \node at (3,1) {1};
    \node at (5,1) {2};
    \node at (7,1) {3};
    \node at (9,1) {4};
    \node at (11,1) {5};
    \node at (13,1) {6};
    \node at (15,1) {7};

    \node[align=left] at (-9.3,1) {Register file:};
\node[align=left, red] at (26.5,-2) {(iii) all $K$ sections\\ in registers};
    \node[align=left] at (-1.5,1) {$\text{T}\ssub{0}$:};

\draw[red, thick, dashed, -latex]
    ([shift={(60:1.1)}]18,-2) arc[start angle=60, end angle=-250, radius=1.1];
\end{scope}

\begin{scope}[xshift=0cm,yshift=-2cm]
    \draw (0,0) rectangle (16,2);
    \node at (1,1) {8};
    \node at (3,1) {9};
    \node at (5,1) {10};
    \node at (7,1) {11};
    \node at (9,1) {12};
    \node at (11,1) {13};
    \node at (13,1) {14};
    \node at (15,1) {15};

    \node[align=left] at (-1.5,1) {$\text{T}\ssub{1}$:};
\end{scope}

\begin{scope}[xshift=0cm,yshift=-4cm]
    \draw (0,0) rectangle (16,2);
    \node at (1,1) {16};
    \node at (3,1) {17};
    \node at (5,1) {18};
    \node at (7,1) {19};
    \node at (9,1) {20};
    \node at (11,1) {21};
    \node at (13,1) {22};
    \node at (15,1) {23};

    \node[align=left] at (-1.5,1) {$\text{T}\ssub{2}$:};
\end{scope}

\begin{scope}[xshift=0cm,yshift=-6cm]
    \draw (0,0) rectangle (16,2);
    \node at (1,1) {24};
    \node at (3,1) {25};
    \node at (5,1) {26};
    \node at (7,1) {27};
    \node at (9,1) {28};
    \node at (11,1) {29};
    \node at (13,1) {30};
    \node at (15,1) {31};

    \node[align=left] at (-1.5,1) {$\text{T}\ssub{3}$:};
\end{scope}

\begin{scope}[xshift=0cm]
    \draw[red,thick] (0,-6) rectangle (2,2);
    \node at (1,1) {};
    \node at (1,-1) {};
    \node at (1,-3) {};
    \node at (1,-5) {};
\end{scope}

\begin{scope}[xshift=2cm]
    \draw[red,thick] (0,-6) rectangle (2,2);
    \node at (1,1) {};
    \node at (1,-1) {};
    \node at (1,-3) {};
    \node at (1,-5) {};
\end{scope}

\draw[-latex, red, thick] (1,-6) -- (1,-11);
\draw[-latex, red, thick] (3,-6) -- (3,-11);
\node[right, red] at (4.0,-8.5) {(iv) row write-back};

\end{scope}

\begin{scope}[yshift=-32cm]

\begin{scope}[xshift=0cm]
    \draw (0,-6) rectangle (2,2);
    \node at (1,1) {0};
    \node at (1,-1) {8};
    \node at (1,-3) {16};
    \node at (1,-5) {24};
\end{scope}

\begin{scope}[xshift=2cm]
    \draw (0,-6) rectangle (2,2);
    \node at (1,1) {1};
    \node at (1,-1) {9};
    \node at (1,-3) {17};
    \node at (1,-5) {25};
\end{scope}

\begin{scope}[xshift=4cm]
    \draw (0,-6) rectangle (2,2);
    \node at (1,1) {2};
    \node at (1,-1) {10};
    \node at (1,-3) {18};
    \node at (1,-5) {26};
\end{scope}

\begin{scope}[xshift=6cm]
    \draw (0,-6) rectangle (2,2);
    \node at (1,1) {3};
    \node at (1,-1) {11};
    \node at (1,-3) {19};
    \node at (1,-5) {27};
\end{scope}

\begin{scope}[xshift=8cm]
    \draw (0,-6) rectangle (2,2);
    \node at (1,1) {4};
    \node at (1,-1) {12};
    \node at (1,-3) {20};
    \node at (1,-5) {28};
\end{scope}

\begin{scope}[xshift=10cm]
    \draw (0,-6) rectangle (2,2);
    \node at (1,1) {5};
    \node at (1,-1) {13};
    \node at (1,-3) {21};
    \node at (1,-5) {29};
\end{scope}

\begin{scope}[xshift=12cm]
    \draw (0,-6) rectangle (2,2);
    \node at (1,1) {6};
    \node at (1,-1) {14};
    \node at (1,-3) {22};
    \node at (1,-5) {30};
\end{scope}

\begin{scope}[xshift=14cm]
    \draw (0,-6) rectangle (2,2);
    \node at (1,1) {7};
    \node at (1,-1) {15};
    \node at (1,-3) {23};
    \node at (1,-5) {31};
\end{scope}

\draw (16,0) rectangle (18,2);
\draw (16,-2) rectangle (18,0);
\draw (16,-4) rectangle (18,-2);
\draw (16,-6) rectangle (18,-4);

\begin{scope}[xshift=0cm]
    \draw[red, thick] (0,0) rectangle (8,2);
    \node at (1,1) {};
    \node at (3,1) {};
    \node at (5,1) {};
    \node at (7,1) {};
\end{scope}

\begin{scope}[xshift=8cm]
    \draw[red, thick] (0,0) rectangle (8,2);
    \node at (1,1) {};
    \node at (3,1) {};
    \node at (5,1) {};
    \node at (7,1) {};
\end{scope}

\node[align=left] at (-8,1) {Shared memory:};

\draw[-latex, red, thick] (4,0) -- (4,-10);
\draw[-latex, red, thick] (12,0) -- (12,-10);
\node[right, red] at (13.0,-8.3) {(v) coalesced store};

\end{scope}

\begin{scope}[yshift=-44cm]

\draw (0,0) rectangle (8,2);

\node at (1,1) {0};
\node at (3,1) {1};
\node at (5,1) {2};
\node at (7,1) {3};

\begin{scope}[xshift=8cm]
    \draw (0,0) rectangle (8,2);
    \node at (1,1) {4};
    \node at (3,1) {5};
    \node at (5,1) {6};
    \node at (7,1) {7};
\end{scope}

\begin{scope}[xshift=16cm]
    \draw (0,0) rectangle (8,2);
    \node at (1,1) {8};
    \node at (3,1) {9};
    \node at (5,1) {10};
    \node at (7,1) {11};
\end{scope}

\begin{scope}[xshift=24cm]
    \draw (0,0) rectangle (8,2);
    \node at (1,1) {12};
    \node at (3,1) {13};
    \node at (5,1) {14};
    \node at (7,1) {15};
\end{scope}

\node[align=left] at (-8,1) {Global memory:};

\node at (34,1) {$\cdots$};

\end{scope}

\end{tikzpicture}
}

\caption{Memory-access pattern at group entry and exit ($L{=}4$
threads, $N{=}8$ blocks shown); steps (i)--(v) are described in
the text.}
\label{fig:staging}
\end{figure}

The multi-block algorithms operate on stride-$N$ permuted data:
thread $i$ holds the $N$ consecutive samples
$x[iN],\ldots,x[iN{+}N{-}1]$, so that at block step $m$ the warp's
$L$ lanes collectively hold the permuted block $\{x[m{+}iN]\}_i$.
Issued directly against global memory, such strided access would
fragment into up to 32 separate transactions, because the memory
system serves contiguous 32-thread transactions only. The
implementation therefore never lets a strided address reach DRAM.
Figure~\ref{fig:staging} shows the pattern at every group boundary:
(i)~the TB loads its $C$ samples contiguously into a shared-memory
array of $L$ rows and $N{+}1$ columns; (ii)~each thread reads its row
into registers, kept free of bank conflicts by the one-word row
padding; (iii)~all $K$ sections execute in registers; (iv)~results
are staged back row-wise; and (v)~committed to global memory
contiguously. With single-precision (4-byte) samples, the buffer
costs
\begin{equation}
S_{\text{buf}} = 4L(N+1) \text{ bytes per TB,}
\label{eq:sbuf}
\end{equation}
which will later be shown to be the limiting resource of the
configuration space (Section~\ref{sec:gpu:config}). Because the
cascade is fused, the staging is paid twice per group---at entry and
exit---regardless of $K$, so the per-section staging cost falls as
$1/K$; this is the same amortization the CPU implementation obtains
from its shuffle-based permutation.

\subsection{Decoupled Lookback for State Propagation}
\label{sec:gpu:lookback}

Groups are independent except for the filter state: section $s$ of
group $c$ needs the last two outputs of section $s$ of group $c{-}1$
as its initial conditions. Left untreated, this dependency chains
every group to its predecessor and the grid degenerates into a serial
pipeline. The decoupled lookback protocol~\cite{Merrill_16}, extended
to higher-order recurrences in~\cite{Maleki_18}, removes the
serialization by exploiting linearity, exactly as in
Section~\ref{sec:bg:prop}: because the initial state enters
additively, a group computes almost all of its output before its
state is known.

\begin{figure}[t]
\centering
\scalebox{0.95}{
\begin{tikzpicture}[
    node distance=0.75cm and 0.75cm,
    biquad/.style={circle, draw, minimum size=0.45cm, inner sep=0pt, font=\tiny},
    arr/.style={->, >=stealth, thin}
]
\definecolor{t1}{RGB}{228,26,28}
\definecolor{t2}{RGB}{55,126,184}
\definecolor{t3}{RGB}{77,175,74}
\definecolor{t4}{RGB}{152,78,163}
\definecolor{t5}{RGB}{255,127,0}

\foreach \r in {0,1,2,3,4} {
  \foreach \g/\c in {0/t1, 1/t2, 2/t3, 3/t4, 4/t5} {
    \node[biquad, fill=\c!40] (n\r\g) at (\g*1.2, -\r*1.6) {};
  }
}

\foreach \r [evaluate=\r as \rn using int(\r+1)] in {0,1,2,3} {
  \begin{scope}[yshift=-\r*1.6cm]

  \draw (-0.45,-0.55) rectangle (5.25,-0.85);

  \foreach \g in {0,1,2,3} {
    \draw[arr, gray!60] (\g*1.2+0.3,-0.7) -- (\g*1.2+0.9,-0.7);
  }

  \foreach \g in {1,2,3,4} {
    \draw (\g*1.2-0.125,-0.85) rectangle (\g*1.2+0.125,-1.10);
  }

  \end{scope}

  \foreach \g in {0,1,2,3,4} {
    \draw[arr] (n\r\g) -- (\g*1.2,{-\r*1.6-0.55});
  }

  \draw[arr] (0,{-\r*1.6-0.85}) -- (n\rn0);

  \foreach \g in {1,2,3,4} {
    \draw[arr] (\g*1.2,{-\r*1.6-1.10}) -- (n\rn\g);
  }
}

\node[right, font=\scriptsize] at (5.45,-0.7) {DRAM};

\node[above=0.1cm of n00, font=\scriptsize] {group 0};
\node[above=0.1cm of n01, font=\scriptsize] {group 1};
\node[above=0.1cm of n04, font=\scriptsize] {group 4};

\node[left=0.15cm of n00, font=\scriptsize] {$F_1$};
\node[left=0.15cm of n10, font=\scriptsize] {$F_2$};
\node[left=0.15cm of n40, font=\scriptsize] {$F_K$};

\draw[arr, thick] (-1.2,0.2) -- (-1.2,-6.6) node[midway, left, font=\scriptsize, rotate=90, anchor=south] {filter order / time};
\draw[arr, thick] (-0.7,1.0) -- (5.1,1.0) node[midway, above, font=\scriptsize] {signal group / thread block};

\end{tikzpicture}
}
\caption{Decoupled lookback execution of the same cascade as
Figure~\ref{fig:wavefront}; one thread block carries one group
(color) through all sections. Local carries are written to, and read
from, a descriptor array in DRAM shared by all thread blocks (gray
arrows); the small rectangles resolve each group's global carry
before the next section.}
\label{fig:lookback}
\end{figure}
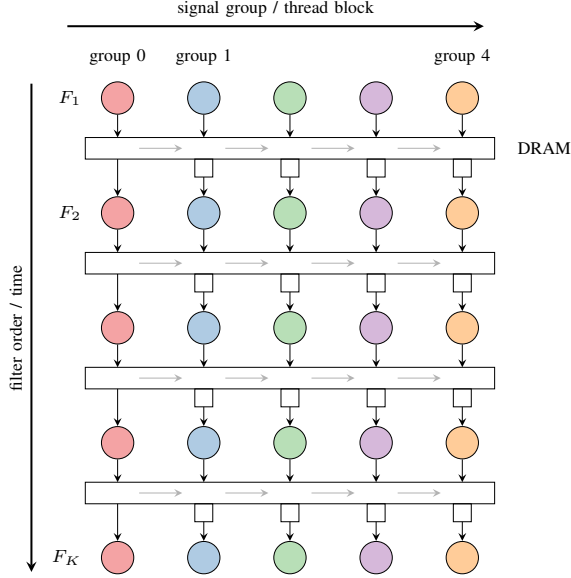

Figure~\ref{fig:lookback} shows the execution. The protocol runs once
per section. For section $s$, the TB first performs the \emph{local
solve}: it filters the section under zero initial conditions, which
is the overwhelming majority of the arithmetic and requires no
communication; the byproduct is the section's \emph{local carry}, its
terminal output pair under zero state. The TB publishes the local
carry to a global descriptor array, issues a memory fence, and sets a
status flag. It then performs the \emph{lookback}: one designated
warp walks backward through the predecessors' descriptors (the gray
arrows in Figure~\ref{fig:lookback}); if predecessor $c{-}j$ has
posted its \emph{global} carry (its resolved terminal state), the
walk stops, and the TB composes that global carry with the local
carries of the intervening groups through the $2 \times 2$ matrix
action of a full group, resolving its own global carry (the small
rectangles in Figure~\ref{fig:lookback}). The lookback distance is
therefore variable: the TB waits only for its most recent
\emph{finished} ancestor, however far back, not for its immediate
predecessor to finish. Finally, the TB applies the state
\emph{correction}---for PH, the homogeneous
solution~\eqref{eq:PH_forward}---publishes the resolved carry, and
proceeds to section $s{+}1$. The protocol's overhead is a few
operations per predecessor inspected against $O(C)$ useful work per
group, and the wait itself is hidden by the SM scheduling other
resident groups.

The CPU and GPU split on this choice for structural reasons. The
wavefront of Section~\ref{sec:cpu} parallelizes across
\emph{sections}: at most $K$ groups are in flight, the state hand-off
is free, FIFO order is preserved, and the cheaper one-sided cyclic
reduction applies. This matches a streaming CPU, where the core count
is close to $K$ and output order matters. The GPU must instead fill
$15 \times 2048 = 30{,}720$ (GTX~1070) or $28 \times 1536 = 43{,}008$
(RTX~3060) resident threads to hide latency; the section axis offers few units
(usually $K \le 8$), so only the group axis is long enough, and
batched processing supplies the required backlog while imposing no
output order. Lookback is therefore necessary on the GPU rather than merely
preferred. Its price is paid in Section~\ref{sec:bg:dccr}'s currency:
publishing the complete terminal state early forces cyclic reduction
into its divide-and-conquer form.

\subsection{The GPU Cost Model}
\label{sec:gpu:model}

Part~I~\cite{Zhai_26} ranked the algorithms by operation count, the right measure
for a processor whose bottleneck is arithmetic latency along a
dependency chain. The GPU prices the same dependency graphs
differently: operations are nearly free, because each warp's
arithmetic latency is hidden by switching to another resident warp,
while every dependency whose producer and consumer sit in different
threads must execute as an explicit exchange. A register shuffle
moves a value between lanes of one warp and is issued and hidden like
arithmetic. A cross-warp exchange is a store to shared memory, a
block-wide barrier, and a load; the barrier stalls every warp of the
TB, and warp switching cannot hide it, because the waiting warps are
the ones that would provide the hiding. A cross-TB exchange is a
global-memory round trip. Following the standard cost models of
communication-avoiding linear algebra~\cite{Ballard_11}, BSP
synchronization~\cite{Valiant_90}, the roofline
bound~\cite{Williams_09}, and latency-hiding
analysis~\cite{Volkov_16,Hong_09}, we model the steady-state time
per sample as
\begin{equation}
t \;\approx\; t_{\text{mem}} + \frac{F}{C} + \beta\,\sigma +
\varepsilon_{\text{hide}}.
\label{eq:model}
\end{equation}

\emph{Memory floor.} Every algorithm reads each input sample once and
writes each output once; intermediates stay on chip and the carries
add a negligible $O(K/C)$ bytes per sample. At single precision this
is $b_s = 8$ bytes of irreducible DRAM traffic per sample, so
\begin{equation}
t_{\text{mem}} = b_s / BW,
\label{eq:tmem}
\end{equation}
giving throughput roofs of $256/8 = 32$~\GSs\ (GTX~1070) and $360/8 =
45$~\GSs\ (RTX~3060). No kernel can exceed its device's roof, and the experimental measurements (Section~\ref{sec:gpu:results})
normalize every plateau by it.

\emph{No arithmetic term.} A cascade performs roughly $6K$ FMAs per
sample (PH; $8K$ for STCR) against the fixed 8 bytes, an arithmetic
intensity of $1.5K$--$2K$ FLOP per byte. The devices' balance points
(peak FLOP/s over bandwidth) sit near 25 and 35 FLOP per byte, so the
workload is memory-bound everywhere and the arithmetic term is
dropped. The measurements test this deletion directly, and locate its
boundary where the margin is thinnest---order 16 on the GTX~1070.

\emph{Fixed per-group overhead.} $F$ collects the constant work a
group performs because it has boundaries: the ticket draw, reading
and injecting carries, publishing with fences and flags, and the
entry/exit staging synchronizations. It is independent of $C$, hence
the dilution $F/C$: negligible at the plateau, visible at small batch
sizes.

\emph{Barrier term.} $\sigma$ counts TB barriers per sample---an
algorithm and configuration property---and $\beta$ is the machine's
cost per barrier: an architecture property. The separation lets the
measurements attribute a gap to the algorithm on one device and to
the architecture between devices.

\emph{Exposed latency.} $\varepsilon_{\text{hide}}$ is the latency
that resident parallelism fails to cover~\cite{Volkov_16}. Keeping a
pipeline of latency $\lambda$ cycles busy at an issue rate of $I$
warp instructions per cycle requires $I\lambda$ instructions in
flight per SM. The supply has two factors. Across warps, every
resident warp (resident TBs $\times$ warps per TB) is an independent
instruction stream. Within a warp, instructions issue back to back
until one needs a result still in the pipeline, so the number a warp
keeps in flight---its instruction-level parallelism (ILP)---is set
by the dependency structure of its inner loop: how many instructions
the loop issues between consecutive dependent ones. ILP is therefore
a property of the kernel's code, and it differs across kernels. A single floor follows from the least-parallel loop in the suite, the
per-thread particular solution~\eqref{eq:particular_recurrence}: of
its three instructions per iteration---two FMAs and one load---only
one FMA depends on the preceding iteration, so three instructions
issue per dependent step and $\text{ILP} \approx 3$. Every other
loop offers at least this much. Setting supply equal to
demand, the minimum resident warps per SM are
\begin{equation}
W_{\min} = \lceil I\,\lambda/\text{ILP} \rceil,
\label{eq:wmin}
\end{equation}
and with $I = 4$ and a dependent FMA latency of $\lambda = 6$ cycles
on Pascal~\cite{Jia_19} and $4$ cycles on
Ampere~\cite{Abdelkhalik_22}, \eqref{eq:wmin} gives $W_{\min} = 8$
and 6 warps (Table~\ref{tab:devices}). Memory and synchronization
latency add further demand, so $W_{\min}$ is a floor: configurations
at or below it should underperform, which the \cfg{32}{128} row of
Table~\ref{tab:configs} confirms.

The remainder of the section is organized around this model.
Section~\ref{sec:gpu:config} applies \eqref{eq:model}, together with
the shared-memory budget, to reduce the configuration space to the
two parameters $L$ and $N$ and four candidate configurations, with
\eqref{eq:wmin} flagging one of them as hiding-starved.
Section~\ref{sec:gpu:algos} prices each algorithm's exchange
schedule and barrier count in the model's terms
(Table~\ref{tab:bills}), yielding testable predictions: where PH and
STCR coincide, where their gap opens, and when DTCR's trade of
barriers for warps pays. Section~\ref{sec:gpu:results} then confronts
these predictions with measurement; no term of \eqref{eq:model} is
fitted to data---the model supplies signs, floors, and ratios, and
the experiments check them.

\subsection[Configuration: The Roles of L and N]{Configuration: The Roles of $L$ and $N$}
\label{sec:gpu:config}

\begin{table}[!t]
\caption{Cascade configurations against the per-SM budgets of
Table~\ref{tab:devices}; device pairs read GTX~1070\,/\,RTX~3060.
Shared memory limits every TBs/SM entry except \cfg{32}{32} on the
RTX~3060, where the TB-slot cap limits first. Regs/thread is the
ptxas-measured count, shared by PH and STCR; DTCR uses about half.
$^\dagger$below the hiding floor of Table~\ref{tab:devices}.}
\label{tab:configs}
\centering
\footnotesize
\setlength{\tabcolsep}{3pt}
\begin{tabular}{@{}lccccc@{}}
\toprule
$L{\times}N$ & $S_{\mathrm{buf}}$ & TBs/SM & warps/SM & regs/thread &
\shortstack{$\sigma_{\mathrm{sec}}$\\[-1pt] PH/STCR/DTCR} \\
\midrule
$32{\times}32$  & 4.2\,kB  & 23 / 16 & 23 / 16 & 72  & 0 / 0 / 2 \\
$32{\times}64$  & 8.3\,kB  & 11 / 12 & 11 / 12 & 107 & 0 / 0 / 2 \\
$32{\times}128$ & 16.5\,kB & 5 / 6   & 5$^\dagger$ / 6 & 168 & 0 / 0 / 2 \\
$64{\times}64$  & 16.6\,kB & 5 / 6   & 10 / 12 & 105 & 1 / 2 / 2 \\
\bottomrule
\end{tabular}
\end{table}

The multi-block algorithms expose two tuning parameters: $L$ threads
per group and $N$ samples per thread, with group size $C = NL$. Their
roles on the GPU differ sharply from the CPU, where $L$ was the SIMD
width (amortizing instructions) and $N$ set the function size.

\emph{The role of $L$.} Solving a length-$L$ coupled recurrence
across $L$ threads requires $\log_2 L$ pairwise exchange rounds at
distances $1, 2, \ldots, L/2$; this is a lower bound on parallel
evaluation~\cite{Hyafil_77}, not an algorithmic choice. PH traverses
this exchange schedule once per section, cyclic reduction twice.
Every round at distance below 32 is an intra-warp shuffle, priced at
zero by \eqref{eq:model}; any round at distance 32 or more crosses
warps and becomes a barrier-synchronized exchange. Under SIMT, each
sample owns its own lane, so the per-sample operation count is
invariant in $L$: unlike the CPU, $L$ buys no arithmetic. Below 32,
lanes idle, so $L = 32$ is the smallest full-utilization width---all
rounds shuffles, zero barriers. Each doubling beyond 32 adds one
barrier-priced round but enlarges the group, diluting $F/C$ without
spending warps.

\emph{The role of $N$.} Raising $N$ dilutes $F/C$ and increases
per-thread ILP, and never touches the exchange schedule. Its costs:
at fixed batch size the group count $n/C$ falls, starving the grid;
and the staging buffer~\eqref{eq:sbuf} grows linearly in $N$. Shared
memory is the hard limit on the configuration. When it is the limiting
budget item---and Table~\ref{tab:configs} shows it is---the resident TBs
per SM are $S_{\text{SM}}/(4L(N{+}1))$ and each TB holds $L/32$
warps, so the factors of $L$ cancel:
\begin{equation}
\text{warps per SM} \;\approx\; \frac{S_{\text{SM}}}{128\,(N+1)},
\label{eq:warps}
\end{equation}
a function of $N$ alone. Raising $N$ costs resident warps roughly
one-for-one; raising $L$ merely repackages the same warps into fewer,
larger TBs.

The tested configurations follow from one intrinsic constraint and
three hardware budgets. The proposed multi-block filtering  requires 
$N = kL$ for integer $k$. Beyond this, the 48-kB per-TB
shared-memory cap excludes $L \ge 128$; warp utilization excludes
$L < 32$; and \eqref{eq:warps} excludes large $N$ (at \cfg{32}{256}
residency collapses to 2--3 TBs per SM). The four survivors are
\cfg{32}{32}, \cfg{32}{64}, \cfg{32}{128}, and \cfg{64}{64}
(Table~\ref{tab:configs}). The last two reach the same $C = 4096$ by
different routes---\cfg{32}{128} pays in resident warps, \cfg{64}{64}
pays one cross-warp barrier round---so their head-to-head is a
controlled comparison of the two currencies. Registers never bind
before shared memory: the cascade footprint is set by the
configuration ($N$ resident samples plus a fixed overhead) and does
not grow with the filter order, since the $K$ sections reuse the
same registers.

\subsection{The Four Algorithms}
\label{sec:gpu:algos}

\begin{table}[!t]
\caption{Communication cost per section per group (per sample for
PLR). Handoff: how section $s$'s output reaches section $s{+}1$.}
\label{tab:bills}
\centering
\footnotesize
\setlength{\tabcolsep}{4pt}
\begin{tabular}{@{}lllll@{}}
\toprule
 & PH & STCR & DTCR & PLR \\
\midrule
FMAs / sample & $\sim 6$ & $\sim 8$ & $\sim 4{\times}2$ & $k\log_2 B$ \\
Exchange rounds & $\log_2 L$ & $2\log_2 L$ & $2\log_2 L$ split & $\log_2 B$ \\
Barriers, $L{\le}32$ & 0 & 0 & 2 & $\log_2\!\frac{B}{32}$+stg. \\
Barriers, \cfg{64}{64} & 1 & 2 & 2 & --- \\
Section handoff & registers & registers & shared mem. & none \\
\bottomrule
\end{tabular}
\end{table}

Each algorithm is described by what changes relative to
Section~\ref{sec:background}; Table~\ref{tab:bills} collects the
costs.

\emph{PH factorization.} PH transfers unchanged: $L$ threads, each
holding $N$ permuted samples. The particular
solution~\eqref{eq:particular_recurrence} is per-thread arithmetic
with no exchange at all---the $\mathcal{O}(N)$ dependency chain that
limits it on the CPU is hidden by residency here. The only
inter-thread communication is the terminal recursive doubling, one
traversal of the exchange schedule ($\log_2 L$ rounds,
Sklansky-style). At $L = 32$ every round is a shuffle and the barrier
count is zero; at $L = 64$, exactly one round crosses warps---the
last, whose exchange distance of $32$ exceeds the warp width, while
the earlier rounds at distances $1, 2, \ldots, 16$ remain in-warp.

\emph{Single-thread-group cyclic reduction (STCR).} The lookback
protocol needs both terminal blocks of a group published before the
state-dependent remainder, so the GPU implementation must use the
divide-and-conquer formulation of Section~\ref{sec:bg:dccr}. A
\emph{thread group} is the set of $L$ threads that jointly holds one
signal block group; STCR uses a single thread group, which reduces
both parities in sequence. Against PH this costs roughly 8 FMAs per
sample instead of 6, and twice the exchange rounds. On the CPU this
operation gap decides the ranking; the model prices it at zero
wherever the rounds stay in-warp, predicting that PH and STCR
coincide in every $L = 32$ configuration.

\emph{Double-thread-group cyclic reduction (DTCR).} DTCR has no CPU
counterpart. It assigns the even- and odd-parity reductions to two
thread groups of $L$ threads each, running concurrently within one
TB.
This doubles the resident warps per group at the same staging
footprint---additional hiding supply for $\varepsilon_{\text{hide}}$.
The cost is the cascade handoff: the two thread groups hold disjoint
halves of the samples, so every section boundary requires a
shared-memory exchange (two barriers per section, versus the register
handoff of PH and STCR). The model therefore predicts a conditional
outcome: DTCR wins where hiding is scarce and loses where the added
barrier cost exceeds the hiding it buys, with the boundary set by the
device's cost of barrier $\beta$.

\emph{Parallel linear recurrence (PLR).} PLR~\cite{Maleki_18} is the
highest-throughput published general linear-recurrence engine and
serves as the literature baseline, in its native realization: the
$K$ sections are combined offline via the $z$-transform into one
direct-form recurrence of order $2K$, computed by hierarchical chunk
merging with precomputed corrections; the published kernels are
transplanted unchanged except for the shared descriptor flags and
the device geometry. The direct form pays for its generality twice
as the order grows. Its per-thread carry footprint grows with the
order: block sizes must halve per order doubling to stay ahead of
the register budget, and at order 16 no spill-free configuration
exists on either device. And its coefficients degrade numerically with the order, leading to numerical failure at order 16 (Section~\ref{sec:gpu:results}).

\subsection{Measurement and Results}
\label{sec:gpu:results}

All experiments use Butterworth low-pass designs of order $2K$, $K
\in \{1, 2, 4, 8\}$, cutoff $0.2\pi$, designed in second-order
sections and $b_0$-normalized in float32; for PLR the same design is
combined offline into the equivalent direct form, so the two
realizations share the transfer function. Batch sizes span
$2^{16}$ to $2^{25}$ samples. Every timed cell runs an accuracy gate:
relative error below $10^{-4}$ of the float64 reference peak. The
cascade algorithms pass at every order on both devices (errors of
order $10^{-7}$). Direct-form PLR passes at orders 2 and 4, is
rounding-limited at order 8 (the float32 direct-form coefficients
alone exceed the gate; its throughput is reported under annotation),
and fails entirely at order 16, where rounding pushes poles outside
the unit circle and the output diverges---the original argument for
the cascaded realization. Each cell launches its kernel 4000 times as 20 batches of 200 after
a 1000-launch warm-up, with a \emph{keep-alive} kernel holding the boost
clock and rotating input buffers defeating the L2 cache; the reported
statistic is the 20th-percentile batch mean, reproducible across
sessions within $\pm 3\%$. The \emph{plateau} of a batch curve is the mean of its
$2^{24}$ and $2^{25}$ points.

\begin{figure*}[!t]
\centering
\includegraphics[width=\textwidth]{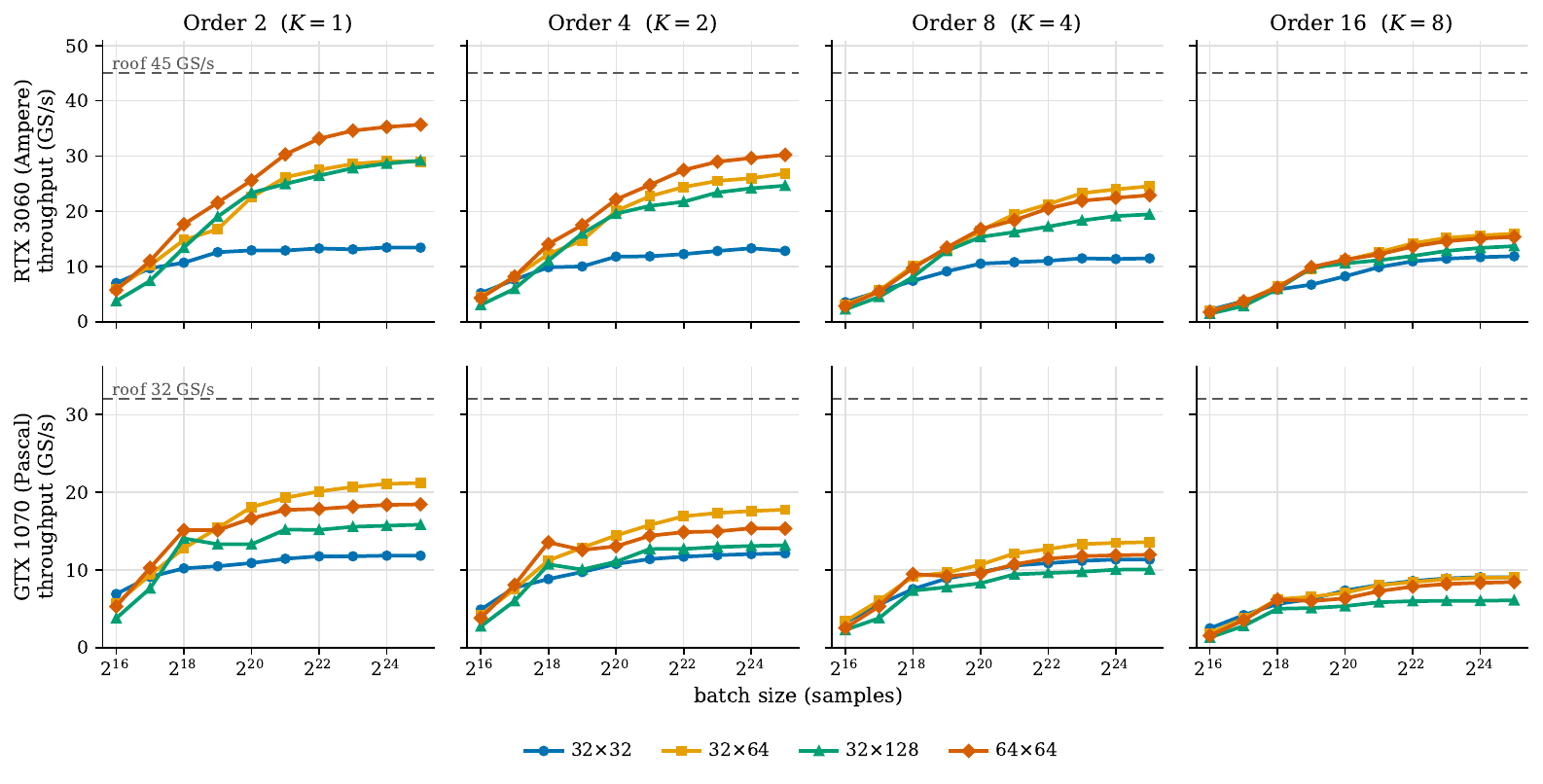}
\caption{PH throughput across the four $L{\times}N$ configurations,
against each device's memory roof~\eqref{eq:tmem} (dashed).}
\label{fig:ph_configs}
\end{figure*}

\emph{PH across the configurations.}
Figure~\ref{fig:ph_configs} shows PH across the four configurations and all orders. Below
saturation, throughput is governed by the group count $n/C$: at
$2^{16}$ one of the two small-group configurations leads every cell,
and neither $C = 4096$ configuration ever does, because more groups
fill the SMs at smaller batches. Past saturation the curves rise
slowly as the per-launch costs amortize; on the 15-SM GTX~1070 the
large-group configurations also show a local peak at $2^{18}$,
where the group count fits a single residency wave, followed by a
dip when a second, partial wave appears. At the plateau, the
configuration analysis of Section~\ref{sec:gpu:config} is visible in
full: the order-2 leaders approach the memory roof, and
\cfg{32}{128}, the configuration at the hiding floor of
Table~\ref{tab:configs}, trails on both devices. As the order grows,
every curve falls away from the roof---the per-section cost grows
while the memory floor does not---but at sharply different rates:
\cfg{64}{64}, with the largest per-section barrier cost, falls
fastest, from 35.5 to 15.2~\GSs\ on the RTX~3060 between orders 2
and 16, while \cfg{32}{32} barely moves, from 13.4 to 11.8~\GSs.
The latter is dominated by the fixed per-group cost $F/C$ rather
than by per-section work, so additional sections are absorbed by
stall slots that were idle anyway: filter order is nearly free
exactly where the fixed overhead dominates.

\begin{figure}[!t]
\centering
\includegraphics[width=\columnwidth]{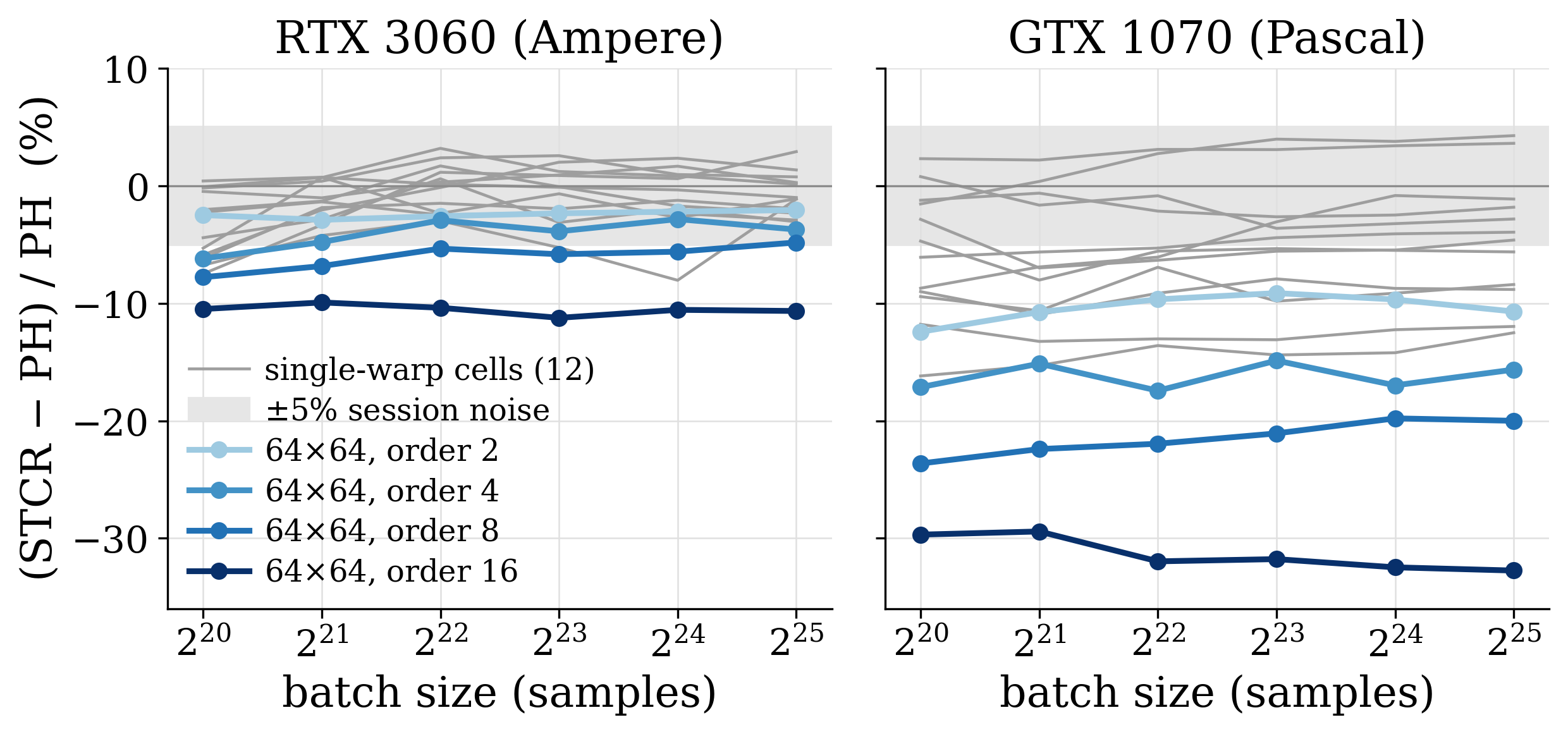}
\caption{(STCR$-$PH)/PH at the plateau. Gray: the twelve single-warp
cells (three $L{=}32$ configurations $\times$ four orders), inside
the $\pm5\%$ session-noise band---the operation gap has no measurable
price. Color: the \cfg{64}{64} cells, where one exchange round
crosses warps; the gap grows with the order and is priced three to
five times higher on Pascal. The gray cells leaving the band on the
GTX~1070 at orders 8 and 16 mark the balance-point boundary discussed
in the text.}
\label{fig:gap}
\end{figure}

\emph{Operations are free.} PH and STCR carry a factor-of-1.5
operation gap into every $L = 32$ configuration, and the gap buys
nothing: at the plateau, the twelve single-warp cells (three
configurations $\times$ four orders) coincide inside the $\pm 5\%$
session-noise band (Figure~\ref{fig:gap}), with both signs. On the
RTX~3060 this holds at every order (worst case $-4.6\%$). The
GTX~1070 locates the boundary of the claim: the band holds through
order 4, then opens to $-9\%$ at order 8 and $-13\%$ at order 16. The
model itself predicts the opening: at order 16 the kernels perform
12--16 FLOP per byte against the 1070's balance point near 25, so the
premise that priced arithmetic at zero is expiring---first on the
device with the lower balance point.

\emph{The barrier price.} Where the exchange schedule crosses warps
(\cfg{64}{64}), STCR's doubled traversal becomes visible, and the gap
grows with $K$: from $-2.1\%$ at order 2 to $-10.6\%$ at order 16 on
the RTX~3060, and from $-10.2\%$ to $-32.6\%$ on the GTX~1070
(Figure~\ref{fig:gap}). The structural gap is the same on both
devices; the older architecture prices it three to five times higher.
In the terms of \eqref{eq:model}, $\sigma$ is the algorithm's
property and $\beta$ the device's.

\begin{figure}[!t]
\centering
\includegraphics[width=\columnwidth]{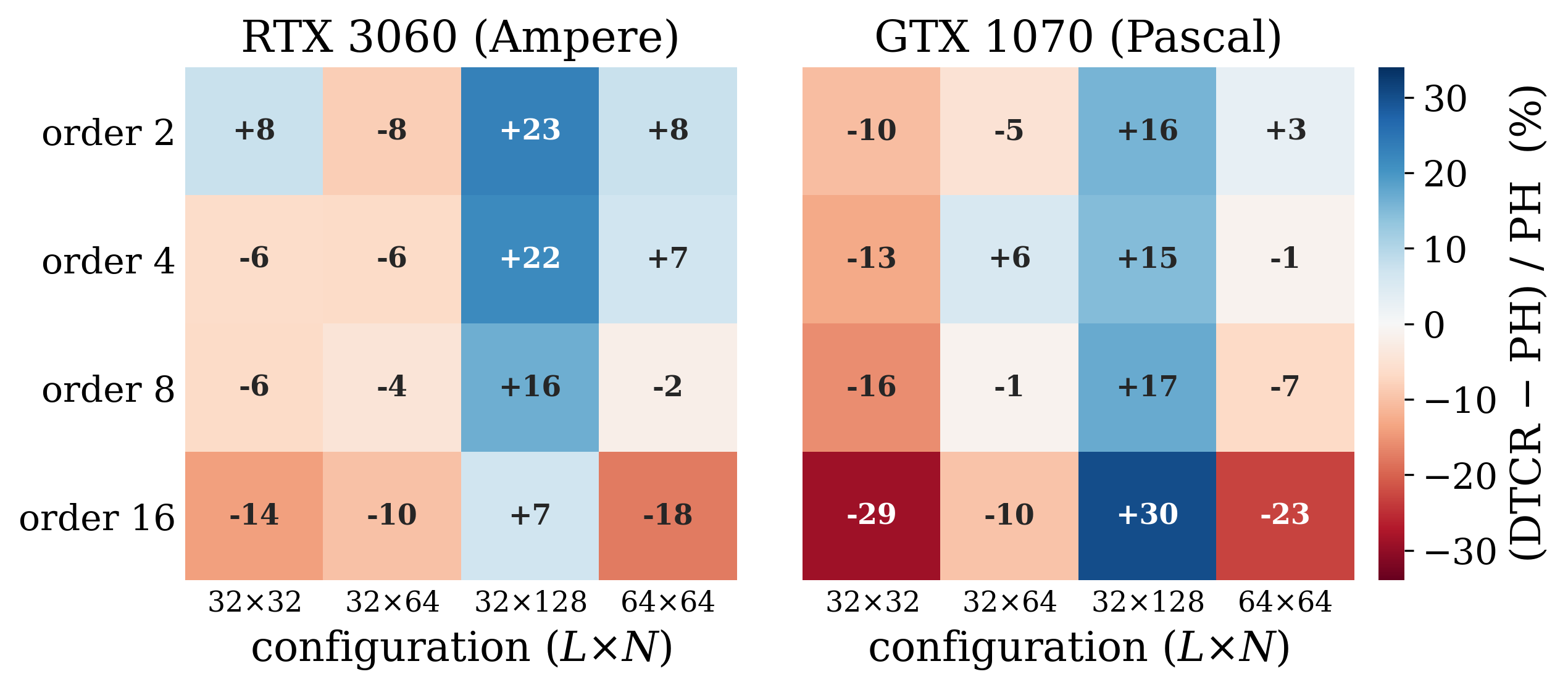}
\caption{DTCR against PH at the plateau, per order and $L{\times}N$
configuration; blue: DTCR wins. The \cfg{32}{128} column---the
hiding-starved configuration---is a win on both devices, the columns
with the most hiding headroom are losses, and cells flipping sign
between the two devices are the barrier price $\beta$ repricing the
same trade.}
\label{fig:dtcr}
\end{figure}

\begin{figure}[!t]
\centering
\includegraphics[width=\columnwidth]{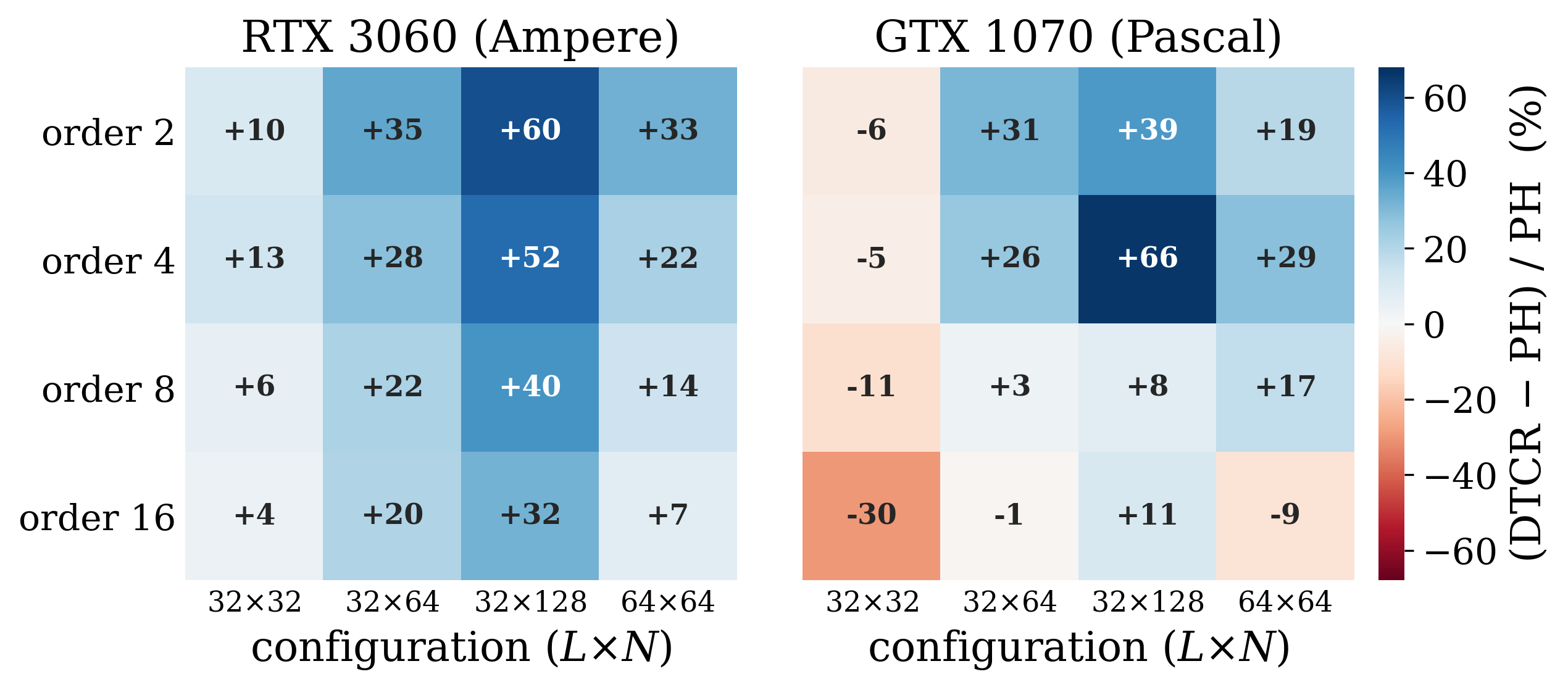}
\caption{The same map at the smallest batch ($2^{16}$): with too few
groups resident to hide latency, DTCR's doubled warps win almost
unconditionally---every cell on the RTX~3060, and the losses on the
GTX~1070 confined to the \cfg{32}{32} column and order 16.}
\label{fig:dtcr16}
\end{figure}

\emph{DTCR's conditional trade.} The DTCR-versus-PH map at the
plateau (Figure~\ref{fig:dtcr}) follows \eqref{eq:model} on both
devices. DTCR wins the entire \cfg{32}{128} column---the
hiding-starved configuration---at every order: $+7\%$ to $+23\%$ on
the RTX~3060 and $+15\%$ to $+30\%$ on the GTX~1070. It never wins
where hiding is most ample (\cfg{32}{64} on the RTX~3060, \cfg{32}{32}
on the GTX~1070), because the added warps hide nothing while the two
barriers per section still cost. Between these poles the boundary
moves with the device, including sign flips (\cfg{32}{32} at order 2:
$+8\%$ on Ampere, $-10\%$ on Pascal). On the RTX~3060 every column
degrades toward order 16 as the per-section barrier cost compounds;
on the GTX~1070 the hiding-starved \cfg{32}{128} column instead
widens in DTCR's favor, to $+30\%$ at order 16, where PH's exposed
latency deepens faster than DTCR's barrier bill. Below
saturation the trade is no longer conditional
(Figure~\ref{fig:dtcr16}): at $2^{16}$, with too few resident groups
to hide latency, DTCR's doubled warps supply the missing hiding and
it wins every cell on the RTX~3060 ($+4\%$ to $+60\%$) and ten of
sixteen on the GTX~1070. DTCR is the algorithm of the latency-bound
regime; the throughput-bound regime prices its barriers.

\emph{Two routes to $C = 4096$, and the best configurations.} In the
controlled comparison of Section~\ref{sec:gpu:config}
(Figure~\ref{fig:ph_configs}), the $L$-route wins on both devices: \cfg{64}{64} beats \cfg{32}{128} by $+22\%$
(RTX~3060) and $+17\%$ (GTX~1070) for PH at order 2, and still by
$+7\%$ and $+4\%$ for DTCR. One cross-warp barrier round costs less
than half the resident warps. Whether the barrier is worth paying at
all, however, depends on the device: against the zero-barrier
\cfg{32}{64}, \cfg{64}{64} gains $+22\%$ on the RTX~3060 and loses
$-13\%$ on the GTX~1070. Accordingly, on the GTX~1070 the
zero-barrier \cfg{32}{64} is the best configuration in eleven of
twelve algorithm--order cells, while on the RTX~3060 \cfg{64}{64}
leads every order-2 and order-4 cell. The best cells are DTCR
\cfg{64}{64} at 38.2~\GSs\ on the RTX~3060 (85\% of its roof) and PH
\cfg{32}{64} at 21.1~\GSs\ on the GTX~1070 (66\%). The best
configuration is a property of the device, not of the algorithm.

\begin{figure}[!t]
\centering
\includegraphics[width=0.92\columnwidth]{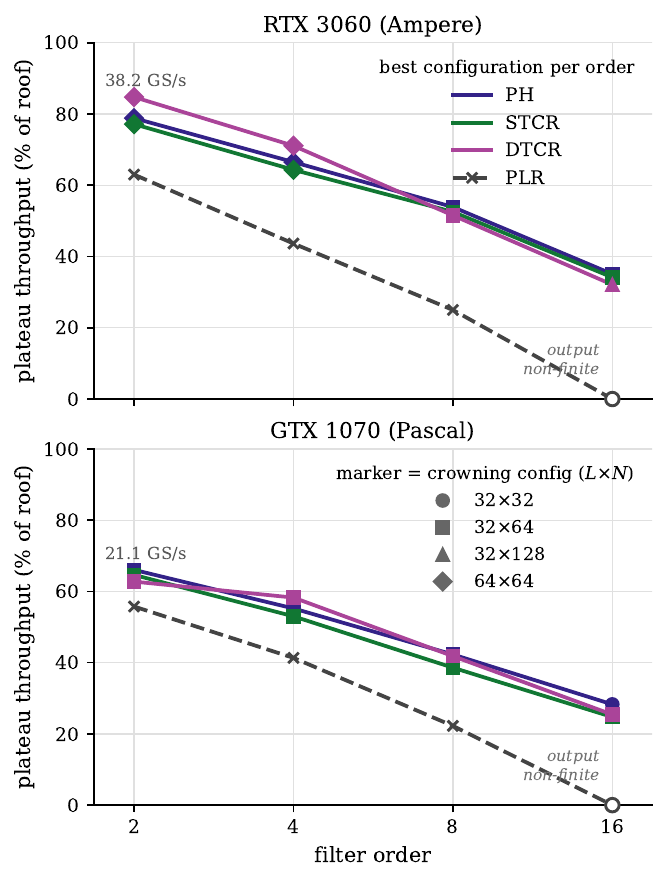}
\caption{The four algorithms at their best configuration per order
(marker: the $L{\times}N$ configuration), as a fraction of each
device's roof \eqref{eq:tmem}. The hollow marker at order 16 is zero
valid PLR throughput: the direct-form output is non-finite there.}
\label{fig:best}
\end{figure}

\emph{PLR.} Figure~\ref{fig:best} compares the four algorithms at
their best configuration per order, as a fraction of each device's
roof. At the order-2 anchor---where cascade and direct form realize
the same recurrence and PLR runs exactly as published---PLR reaches
63\% of the roof on the RTX~3060 and 56\% on the GTX~1070, behind
every cascade leader. The deficit widens with order, tracking its
$k\log_2 B$ work law quantitatively: the predicted order-8/order-4
throughput ratio is 0.57 on the RTX~3060 (measured 0.573) and 0.5625
on the GTX~1070 (measured 0.538). At order 16 PLR contributes no
valid output at any geometry. The order-2 anchor also bounds the
hypothetical cascade of PLR stages without a further measurement: it
inherits at best that per-stage throughput and then adds $K$
inter-section handoffs through global memory, so the single-kernel
cascade algorithms beat it at every order.

\emph{Cross-architecture summary.} Raw throughput conflates algorithm
quality with device hardware: the two devices differ by $1.41\times$
in bandwidth but by $1.7$--$1.8\times$ in measured leaders.
Normalized by each device's roof, the order-2 leaders read 85\%
(Ampere) against 66\% (Pascal). 

The dimensionless structure
transfers: the PH--STCR coincidence inside the noise band, the growth
of the barrier gap with $K$, the sign map of DTCR's trade, and PLR's
order ratios all reproduce on both devices. What does not transfer is
attributable to three named parameters: the barrier price $\beta$
falls by a factor of three to five from Pascal to Ampere, moving the
best configuration from zero-barrier to barrier-tolerant. The hiding
budget grows, visible in every roof fraction. And the order-16
boundary of the free-operations claim belongs to the balance point.
The dependency graphs of Part~I are invariant; the two architectures
price them differently, and this section has measured the prices.

\section{Conclusion}
\label{sec:conclusion}
 
\noindent This paper extended the theoretical multi-block filtering framework \cite{Zhai_26} from a single SIMD core to the two architectures where modern throughput resides: multi-core CPUs and GPUs. The obstacle to that
extension is the inter-group state dependency---the terminal outputs
of each signal block group are the initial conditions of the
next---and its resolution is superposition: each group's output
splits into a zero-state response, computable before the group's
state is known, and a homogeneous correction applied when the state
arrives. Integrating this decomposition into the multi-block
algorithms, and reformulating cyclic reduction in a
divide-and-conquer form that exposes both terminal blocks before
back substitution, allows groups to execute concurrently on either
architecture.
 
The two implementations pair the two deployment scenarios with
opposite treatments of the same dependency. For real-time streaming,
a wavefront pipeline realized with TBB flow graphs parallelizes
across cascade sections, preserving FIFO order and requiring no
state protocol at all; a two-level granularity strategy---function
size within a core, grain size across the task system---yields
$3.95\times$ scaling on the six performance cores of a heterogeneous
Meteor Lake processor, about \SI{2.4}{\giga\samples\per\second} for
a 16th-order filter, and the measurements show that adding
efficiency cores degrades the pipeline rather than extending it. For
batched processing, a single-kernel GPU implementation carries each
group through the entire cascade in registers and parallelizes
across groups with a per-section decoupled lookback protocol. A
communication-based cost model---memory roof, barrier price, and
latency-hiding floor---reduces configuration tuning to the two
parameters $L$ and $N$ and predicts the measured behavior: operation
counts are free while cross-warp barriers are not, the
double-thread-group variant of cyclic reduction wins exactly where
latency hiding is scarce, and the best kernels reach
\SI{38.2}{\giga\samples\per\second} on an RTX~3060, 85\% of the
device's memory-bandwidth roof; for the 16th-order filter, the best
kernel sustains \SI{15.8}{\giga\samples\per\second}, about
$6.5\times$ the multi-core CPU rate. Against the strongest published GPU
recurrence engine, evaluated in its native direct form, the cascade
algorithms are faster at every filter order and remain numerically
valid at order 16, where the direct form produces no usable output.
Normalized by each device's memory roof, the qualitative structure
of the results transfers across two GPU generations, with the
differences attributable to three named hardware parameters: the
barrier price, the latency-hiding budget, and the arithmetic balance
point. All GPU kernels presented in this paper are released as an
open-source CUDA library, along with the raw measurement data,
available at
\url{https://github.com/Haotian-RA/cuda_recursive_filtering}.
 
Together with Part~I, these results establish cascaded second-order
recursive filtering---long considered inherently sequential---as a
first-class workload on every tier of commodity parallel hardware,
from a single SIMD core to a saturated GPU. Future work will pursue
persistent-kernel execution to bring the GPU design into the
low-latency streaming regime, extension of the state propagation
framework to time-varying and adaptive coefficients, and automated
configuration selection driven directly by the cost model.

\bibliographystyle{IEEEtran}

\bibliography{ref}



\end{document}